\documentclass[aps,prx,onecolumn,english,showpacs,superscriptaddress,amssymb,amsfonts]{revtex4-2}
\usepackage{graphicx}
\usepackage{hyperref}
\usepackage{amsmath}
\usepackage{braket}
\usepackage{tikz}
\usepackage{xypic}
\usetikzlibrary{positioning}
\usepackage{soul,xcolor,dsfont}
\usepackage[export]{adjustbox}
\usepackage{placeins}
\usepackage[scr=boondox]{mathalpha}
\usepackage{longtable}
\usepackage{bbm}

\newcommand{\Rom}[1]{\uppercase\expandafter{\romannumeral #1\relax}}
\usepackage{amsthm}

\newtheorem*{assumption*}{\assumptionnumber}
\providecommand{\assumptionnumber}{}


\theoremstyle{definition}

\theoremstyle{theorem}

\theoremstyle{corollary}

\theoremstyle{lemma}

\theoremstyle{Proposition}

\theoremstyle{definition}

\usepackage{amssymb}

\newcommand{\bst}{{\mathcal{T}}}

\makeatletter

\newcommand{\Rmnum}[1]{\expandafter\@slowromancap\romannumeral #1@}
\makeatother
\newcommand{\imth}{\hspace{1pt}\mathrm{i}\hspace{1pt}}

\newcommand{\mbz}{{\mathbb{Z}}}
\newcommand{\bea}{\begin{eqnarray}}
\newcommand{\eea}{\end{eqnarray}}
\newcommand{\bct}{\begin{center}}
\newcommand{\ect}{\end{center}}
\newcommand{\bpm}{\begin{pmatrix}}
\newcommand{\epm}{\end{pmatrix}}
\newcommand{\bal}{\begin{aligned}}
\newcommand{\eal}{\end{aligned}}
\newcommand{\bfr}{\begin{framed}}
\newcommand{\efr}{\end{framed}}

\usepackage{tikz-cd}
\usetikzlibrary{matrix,arrows,backgrounds}

\usepackage{multirow}

\begin{document}
	\title{Pairing Symmetry and Fermion Projective Symmetry Groups}

\author{Xu Yang}
\email{yang.6309@osu.edu}
\author{Shuangyuan Lu}
\thanks{These two authors contributed equally}
\author{Sayak Biswas}
\thanks{These two authors contributed equally}
\author{Mohit Randeria}
\author{Yuan-Ming Lu}
\affiliation{Department of Physics, The Ohio State University, Columbus, OH 43210, USA}
	
	\date{\today}
	
	\begin{abstract}
The Ginzburg-Landau (GL) theory is very successful in describing the pairing symmetry, a fundamental characterization of the broken symmetries in a paired superfluid or superconductor. However, GL theory does not describe fermionic excitations such as Bogoliubov quasiparticles or Andreev bound states that are directly related to topological properties of the superconductor. In this work, we show that the symmetries of the fermionic excitations are captured by a Projective Symmetry Group (PSG), which is a group extension of the bosonic symmetry group in the superconducting state. We further establish a correspondence between the pairing symmetry and the fermion PSG. When the normal and superconducting states share the same spin rotational symmetry, there is a simpler correspondence between the pairing symmetry and the fermion PSG, which we enumerate for all 32 crystalline point groups. We also discuss the general framework for computing PSGs when the spin rotational symmetry is spontaneously broken in the superconducting state. This PSG formalism leads to experimental consequences, and as an example, we show how a given pairing symmetry dictates the classification of topological superconductivity.
    \end{abstract}

 
	\maketitle
 \tableofcontents

\section{Introduction}

One of the most fundamental characterizations of a superconductor or a paired superfluid is the symmetry of its pair wave-function.
The standard way of describing pairing symmetry is in terms of the irreducible representations (irreps) of
the {\it normal state} symmetry group $\mathcal{G}_0$ which constrains the form of the Ginzburg-Landau (GL) free energy functional
\cite{Volovik1985,sigrist1991,annett1990,sigrist2005}.
$\mathcal{G}_0$ can be written as
\begin{equation}
   \mathcal{G}_0 = G_0\times U(1) = \begin{cases}
        X_0 \times SO(3)_{\text{spin}} \times U(1) & \text{Weak SOC}\\
        X_0 \times U(1) & \text{Strong SOC}
    \end{cases}
\end{equation}
where $X_0$ is the crystalline point group, and SOC denotes spin-orbit coupling.
At a second order phase transition, the superconductor spontaneously breaks global charge $U(1)$ symmetry and the system 
condenses into a particular irrep of $G\subseteq G_0$, which is the group of unbroken symmetries in the superconducting phase. For example, $G=X\times SO(3)_{\text{spin}}$ for a singlet superconductor with weak SOC, where $X\subseteq X_0$ is the point group symmetry preserved in the superconductor. In the presence of a strong SOC we have $G=X$ with $X\subseteq X_0$ being the unbroken point group of the superconductor. 

Essentially all of the phonon-mediated superconductors (SCs) exhibit singlet ``$s$-wave" pairing, where the superconducting (SC) state 
transforms according to the trivial representation of $X_n$. But superfluid $^3$He \cite{leggett1975theoretical} and many quantum materials, including
the heavy fermion SCs \cite{pfleiderer2009heavyfermion}, the high $T_c$ cuprates \cite{tsuei2000pairing}, and Sr$_2$RuO$_4$ \cite{mackenzie2003sr2ruo4}, condense into nontrivial irreps.

In this paper, we wish to focus on the relation between pairing symmetry and the symmetry of the Hamiltonian 
describing the {\it fermionic excitations in the superconducting state}.
At the mean field level, one focuses on the Bogoliubov-de Gennes (BdG) Hamiltonian, but the fermionic symmetry analysis
applies equally beyond the BdG framework where one needs to take into account interactions between quasiparticles. 
The approach we develop here will allow us to gain new insights that go beyond the (bosonic) GL theory. 

Examples of questions which this formalism would shed light on include: 
(a) the relation between pairing symmetry and topology, as the
K-theory classification~\cite{ryu2010,Kitaev2009,chiu2016} of non-interacting topological SCs is based on the BdG Hamiltonians, 
(b) how interactions
between quasiparticles for various pairing symmetries impacts the classification of interacting topological SC phases~\cite{fidkowski2010,Wang2014,chiu2016,Wang2020}, 
(c) the relation between pairing symmetry 
and excitations in topological defects such as Majorana zero modes trapped in vortices\cite{Read2000,teo2010,Teo2017}, and 
(d) whether new probes of electronic excitations can provide insight into the pairing symmetry. 
We discussed question (a) in section \ref{sec:consequence} of the manuscript. We will return to other questions in subsequent papers.

Here, we first show how starting with the pairing symmetry, together with the crystalline symmetries that dictate the normal state electronic structure,
we can derive the projective symmetry group (PSG)\cite{Wen2002} for the fermionic excitations in the SC state. Focusing on the cases where the superconductor shares the same spin rotational symmetry as the normal state, we present an exhaustive classification of 
the SC state PSG corresponding to every allowable pairing symmetry for the 32 crystalline point groups with and without SOC.
When confronted with a new superconductor, we would like to use these results in the ``reverse" direction, namely, how can we deduce the possible pairing symmetry, given fermionic properties in the SC state. Mathematically, the map from the pairing symmetry to the SC state PSG is, in general, 
neither injective nor surjective, and thus it cannot be inverted. Nevertheless, we show below that the SC state PSG does constrain to a considerable extent the possible pairing symmetries. We also present numerous examples that serve to illustrate our general results.

To describe the symmetries of the fermionic Hamiltonian we need to
(i) to focus on the {\it superconducting state} symmetry group $G$ as distinct from the {\it normal state} $G_0$ relevant for GL theory,
and (ii) to take into account fermion parity $(-1)^{\hat{F}}$, where ${\hat{F}}$ is the total number of fermions in the system.
Let us discuss each of these points in turn.

On general grounds, the SC state symmetry group $G$ is a subgroup of the normal state $G_0$.
If the irrep into which the GL theory condenses is one-dimensional, then in fact $G = G_0$. 
An example may be useful to illustrate this point.  The $d_{x^2-y^2}$ pairing state in the cuprates transforms according to the $B_{1g}$ irrep of the 
tetragonal symmetry group $D_{4h}$. The pair wavefunction changes sign under a $\pi/2$ rotation, and 
one might naively think that this breaks $C_4$ down to $C_2$.
However, one can compensate for this minus sign by having the fermion operators pick up 
an $e^{\imth\pi/2}$ phase under $C_4$ and thus have the electronic Hamiltonian retain the full symmetry of the normal state. 
We will see a generalization of this at play in the analysis later in section \ref{sec:psg}.

On the other hand, if the irrep has a dimension $> 1$, then one needs to solve the GL equations to find the SC state that minimizes the free energy.
Then the SC state state symmetry is lower than that in the the normal state, and $G$ is a proper subgroup of $G_0$.
For example, $^3$He is a $p$-wave, triplet superfluid, corresponding to the $L = 1, S = 1$ irrep
of the normal state symmetry group $G_0 =SO(3)_{\text{orbital}} \times SO(3)_{\text{spin}}$.
Depending on external parameters various superfluid states are stabilized, and 
in the $B$-phase of $^3$He, for instance, $G_0$ is broken down to $G = SO(3)_{L+S}$~\cite{VollhardtWoelfle1990}. We will discuss a general framework to understand the PSG of fermion excitations in any superconductor in Section \ref{sec:general}, where the superconductor can spontaneously break the normal-state spin rotational symmetry. 

The second point above related to fermion parity may seem trivial: it enforces that 
a Hamiltonian can only have terms with an even number of fermion operators. 
It leads, however, to the important mathematical structure of a projective symmetry group (PSG) $G_f$ acting on the many-body Hilbert space. 
In Section \ref{sec:psg}, we discuss in detail how $G_f$ is built as a central extension of $G$ by the fermion parity group $\mbz_2^{F}$.

The rest of the paper is organized as follows. In Section \ref{sec:PSG-->pairing sym} we show how the fermion PSG $G_f$ can constrain the pairing symmetry of the SC state, applying the framework to all 32 point groups (see Table \ref{table:point_group}) and demonstrating it by a few examples in section \ref{sec:examples}. We further discuss how the PSG determines topological properties of the SC in section \ref{sec:consequence}. While sections \ref{sec:psg}-\ref{sec:PSG-->pairing sym} focus on the cases where the normal state and the SC state shares the same spin rotational symmetries, in section \ref{sec:general extension} we describe a generic theory framework that applies to all superconductors, and further demonstrate its power in the examples of A- and B-phases of superfluid $^3$He in section \ref{sec:helium 3}. Finally we conclude in section \ref{sec:conclusion} with an outlook to future studies.

\section{Characterization of broken symmetries in a superconductor}\label{sec:psg}

\subsection{Projective Symmetry Group and Projective Representation}

Any Hamiltonian must conserve fermion parity $(-1)^{\hat{F}}$  
even if it does not conserve particle number $\hat{F}$, as, for instance, in the presence of pairing.
The fermion symmetry group $G_f$ acting on the many-body Hilbert space of fermions is a projective symmetry group (PSG).
Mathematically, $G_f$ is a central extension of the bosonic symmetry group $G$ in the SC state by the fermion parity group $\mathbb{Z}^F_2 = \left\{(\pm 1)^{\hat{F}}\right\}$.
This may be written as a short exact sequence
\begin{equation}
1\rightarrow \mathbb{Z}^F_2 \rightarrow G_f \rightarrow G \rightarrow 1     
\label{cent_ext}
\end{equation}
where $\mathbb{Z}^F_2$ is in the center of $G_f$. 
Thus fermion parity commutes with all elements of $G_f$ and the quotient group $G_f / \mathbb{Z}^F_2 = G$. 

Let us denote by $\hat{g}$ the operator corresponding to the group element $g \in G$ that acts on Hilbert space. In general it could be unitary or anti-unitary.
The group $G_f$ is then the set $\left\{ (\pm 1)^{\hat{F}}\hat{g} \;\;\; | \; g \in G\right\}$ with the product rule between $(\eta_1)^{\hat{F}}\;\hat{g}$ and $(\eta_2)^{\hat{F}}\;\hat{h}$ (with $\eta_i = \pm 1$) given by
\begin{equation}
   \left[ (\eta_1)^{\hat{F}}\; \hat{g} \right] \left[ (\eta_2)^{\hat{F}}\;\hat{h} \right] =  \left[\eta_1\:\eta_2\;\omega(g,h)\right]^{\hat{F}}\; \widehat{g\,h}
   \label{psg:def}
\end{equation}
$\omega$ called the $2$-cocycle is a function $\omega : G \times G \rightarrow \left\{+1,-1\right\}$ that satisfies 
$\omega(g,h)\;\omega(g h,k) = \omega(g,h k)\;{^g}\omega(h,k)$\footnote{Here we define ${^g}\omega(h,k)=\hat g \omega(h,k) \hat g^{-1}$, which is $\omega(h,k)$ if $g$ is a unitary or $\omega^\ast(h,k)$ if $g$ is an anti-unitary symmetry.}, so that the multiplication is associative, and $\omega(e_G\:,\:e_G) = 1$, 
so that the identity element is well defined. Each inequivalent cocycle furnishes a distinct projective symmetry group. Thus PSGs are 
characterized by classes of inequivalent cocycles $\left[\omega\right]$ which form the second cohomology group $\mathcal{H}^2(G,\mathbb{Z}_2)$.

As an example, consider time reversal symmetry where $G = \mathbb{Z}^T_2 = \left\{ \mathbbm{1},T\right\}$. In this case, $\mathcal{H}^2(\mathbb{Z}_2,\mathbb{Z}_2) = \mathbb{Z}_2$ and there are two PSGs characterized by the two inequivalent cocycles: (1) $\omega(T,T) = 1$ in which case $\hat{T}^2 =1$, 
and (2) $\omega(T,T) = -1$ where $\hat{T}^2 = (-1)^{\hat{F}}$. In the first case $G_f = \mathbb{Z}_2 \times \mathbb{Z}_2$ while in the second  $G_f = \mathbb{Z}_4$. 
Physically, the action of the different PSGs on the even particle number subspace is the same as that of the bosonic group $G$. The distinction appears in how $G_f$ acts on the odd particle number subspace, in particular, the single particle subspace. 

In general, one could have both unitary and anti-unitary symmetries but in this paper we will focus on \textit{unitary} operators $\hat{g} \in G_f$, 
under which the fermion annihilation operator transforms as 
\begin{equation}
    \hat{g} \; \hat{c}_{\mathbf{k} \:\alpha} \; \hat{g}^{-1} = \left[U^g(\mathbf{k})\right]^\dagger_{\alpha \beta} \; \hat{c}_{g\mathbf{k}\:\beta}
    \label{fermion-op-trans}
\end{equation}
where $\mathbf{k}$ is the (crystal) momentum, and the $\alpha$ labels spin, orbital/sublattice/band degrees of freedom (d.o.f.).
Using $(-1)^{\hat{F}} \: \hat{c}_{\mathbf{k}\: \alpha}\: (-1)^{\hat{F}} = -\hat{c}_{\mathbf{k}\: \alpha}$ and eq.~\eqref{psg:def}, we find that 
\begin{equation}
    U^g(h\,\mathbf{k})\:U^h(\mathbf{k}) = \omega(g,h)\: U^{gh}(\mathbf{k}).
    \label{cocycle}
\end{equation}
The $U^g$'s thus form a projective representation of $G$ with coefficients in $\left\{ \pm 1 \right\}$. Equivalently, one can regard $\left\{\pm U_g\;|\;g\in G\right\}$ as a linear representation of $G_f$ with $(-1)^{\hat{F}}$ represented by $-\mathbbm{1}$.

\subsection{Pairing Symmetry and Projective Representations}\label{sec:pairing sym-->PSG}

To be concrete, let us focus on the BdG Hamiltonian
\begin{equation}\label{ham:BdG}
    \hat{H} = \hat{H}_0 + (\hat{H}_{\text{pair}} + \text{h.c.})
\end{equation}
 where 
\begin{equation}\label{ham:kinetic}
    \hat{H}_0 = \sum_{\alpha\, \beta ;\mathbf{k}} \hat{c}^\dagger_{\mathbf{k}\alpha} h_{\alpha \beta}(\mathbf{k}) \hat{c}_{\mathbf{k}\beta}
\end{equation}
is the kinetic energy that describes the normal state electronic dispersion, and 
\begin{equation}\label{ham:pairing}
    \hat{H}_{\text{pair}} = \sum_{\alpha\, \beta; \mathbf{k}} \hat{c}^\dagger_{\mathbf{k}\alpha} \Delta_{\alpha \beta}(\mathbf{k}) \hat{c}^\dagger_{-\mathbf{k}\beta}
\end{equation}
describes the pairing. Fermi statistics dictates that $\Delta_{\alpha\beta }(\mathbf{k}) = - \Delta_{\beta\alpha }(-\mathbf{k})$. 

Initially, we restrict ourselves for simplicity to situations where $SO(3)_{\text{spin}}$ is {\it not} broken spontaneously in the SC state. 
In this case, the SC state symmetry group $G$ is of the form
\begin{equation}
    G = \begin{cases}
        X \times SO(3)_{\text{spin}} & \text{Weak SOC}\\
        X & \text{Strong SOC}
    \end{cases}
    \label{G}
\end{equation}
where $X$ is the point group of crystalline symmetries. In either case the pairing order parameter $\Delta({\bf k})$ forms a 1d linear representation of crystalline point group $X$. Moreover the relevant fermionic PSGs are of the form $G_f = X_f \times SU(2)$ and $G_f = X_f$ for the weak and strong SOC cases respectively where $X_f$ is itself a central extension of $X$ with respect to fermion parity. It is thus sufficient to look at the central extensions of the $X$. 
Later, in Section \ref{sec:general}, we shall present a more general treatment and discuss the case of $^3$He where spin rotation is spontaneously broken in the SC state. In such cases, the fermion symmetry group might have a more complicated form and it is no longer sufficient to look at central extensions of the spatial part alone. 

We now discuss {\it three} different projective representations of $X$ and explore how these are related.
First, we begin with $X^0_f = \left\{ (\pm1)^{\hat{F}}\,\hat{g}_0\;\;\;|\;\; g\, \in \,X   \right\}$ the PSG of $X$ that preserves the kinetic part of the BdG hamiltonian i.e., $\hat{g}_0 \;\hat{H}_0 \; \hat{g}_0^{-1} = \hat{H}_0$. The fermion operators then transform according to the corresponding projective representation 
$U^g_0(\mathbf{k})$, defined by $\hat{g}_0\, c_{\mathbf{k}\, \alpha} \hat{g}_0^{-1}\; = \; [ U^g_0(\mathbf{k}) ]^\dagger _{\alpha \beta} \; \hat{c}_{g\mathbf{k}\, \beta}$, which preserves the normal state band structure
\begin{equation}
    U^g_0(\mathbf{k})\, h(\mathbf{k})\, [U^g_0(\mathbf{k})]^\dagger =  h(g\,\mathbf{k}).
    \label{kinetic}
\end{equation}
We shall call $X_f^0$ the {\it normal state PSG} and denote the corresponding 2-cocycle by $\omega_0$. For systems with weak SOC, crystalline symmetries do not act on the spin degrees of freedom and the PSG is trivial in this case $\omega_0(g,h) \,= \,1$ for any elements $g,h\in X$. In the presence of strong SOC the projective representation in non-trivial with operations like two fold rotations and mirror reflections now squaring to fermion parity, $\omega_0(C_2,C_2) = \omega_0(M,M) = -1$. This becomes evident by looking at the forms of the projective representations in the two cases.
\begin{equation}
    U^g_0(\mathbf{k}) = \begin{cases}
        u^g_{\text{orbital}}(\mathbf{k}) \otimes \mathbbm{1}_{\text{spin}} & \text{weak SOC}\\ u^g_{\text{orbital}}(\mathbf{k}) \otimes e^{i\frac{\theta_g}{2} \hat{n}_g.\Vec{\sigma}} & \text{strong SOC} 
    \end{cases}
\end{equation}
where $\hat n_g$ and $\theta_g$ are the rotation axis and angle associated with crystalline symmetry operation $g\in X$. 

Next, we note that the normal state PSG preserves the pairing term only up to a phase, namely
\begin{equation}
    \hat{g}_0\;\hat H_{\text{pair}}\;\hat {g}_0^{-1}=e^{i\Phi_g}\hat{H}_{\text{pair}}
    \label{pairing-symm}
\end{equation}
The phases $\left\{ e^{i\Phi_g}\;|\; g \in X\right\}$ form a 1D \textit{linear} representation of $X$, which we call the {\it pairing symmetry} $\mathcal{R}_{\text{pair}}$.
The phases $\Phi_g$'s satisfy the equation
\begin{equation}
    \Phi_g + \Phi_h = \Phi_{gh} + 2n\pi \ \ \ (n \in \mathbb{Z}).
    \label{phi-g}
\end{equation}
The pairing matrix $\Delta(\mathbf{k})$ satisfies
\begin{equation}
    U^g_0(\mathbf{k}) \Delta(\mathbf{k}) \left[ U_0^g(-\mathbf{k})\right]^T    \,=\, e^{i\Phi_g}\, \Delta(g\mathbf{k}).
    \label{pairing-symm-2}
\end{equation}

We see from eq.~\eqref{pairing-symm} that the PSG $X^0_f$ that leaves $\hat{H}_0$ invariant, fails to preserve the pairing term. 
However the situation can be fixed as follows.  We modify the transformation of the 
fermions $\hat{g}'\, c_{\mathbf{k} \alpha} \hat{g}'^{-1}\; = \; [  \Tilde{U}(\mathbf{k}) ]^\dagger _{\alpha \beta} \; \hat{c}_{\mathbf{k} \beta}$ with
\begin{equation}
    \Tilde{U}^g(\mathbf{k}) = e^{-i\Phi_g/2} U_0^g(\mathbf{k})
    \label{U-tilde}
\end{equation}
The kinetic part $\hat H_0$, which is invariant under $U(1)$ phase rotations, is preserved by the modified transformations as can be seen from \eqref{kinetic}. 
The new transformations are also symmetries of the pairing term $\hat H_\text{pair}$ as $\Tilde{U}_g(\mathbf{k})$'s lead to eq.~\eqref{pairing-symm-2} without the phase factor $e^{i\Phi_g}$ appearing on the right hand side. 

We thus define {\it SC state PSG} $\Tilde{X}_f$ that preserves the full BdG Hamiltonian by 
\begin{equation}
    \Tilde{X}_f = \left\{ (\pm 1)^{\hat{F}}\;\hat{g}'= (\pm 1)^{\hat{F}}\; e^{-i(\Phi_g/2)\hat{F}}\:\hat{g}\;\;|\;\; g\;\in \; X\right\}
    \label{X-tilde}
\end{equation}
This PSG is characterized by the 2-cocycle $\Tilde{\omega}$.

The last step here is to look at the relation between the normal and the superconducting state PSGs, or equivalently, between their cocycles $\omega_0$ and $\Tilde{\omega}$. The phases $\left\{ e^{-i\Phi_g/2}\;|\; \, g\in X \right\}$ form a 1D \textit{projective} representation of $X$, which we call $\mathcal{R}_\Phi$. This follows from \eqref{phi-g} by observing that $e^{-i\Phi_g/2} e^{-i\Phi_h/2} = (-1)^n\,e^{-i\Phi_{gh}/2}$. From eqn.\eqref{U-tilde} one concludes that the cocycle $\omega_\Phi$ associated with $\mathcal{R}_\Phi$ satisfies
\begin{equation}
    \Tilde{\omega}(g,h) = \omega_\Phi(g,h)\; \omega_0(g,h) 
    \label{wtilde=wphi-w0}
\end{equation}

To summarize, we encountered the following projective representations and their associated cocycles which define the corresponding PSG's:
\begin{subequations}
    \begin{eqnarray}
        \text{Normal state:}\;\;\;\; U_0^g(h\,\mathbf{k})\:U_0^h(\mathbf{k}) &=& \omega_0(g,h)\: U_0^{gh}(\mathbf{k})\\
        \nonumber\\
        \text{$\mathcal{R}_\Phi$ :}\;\;\;\;\;\;\;\;\; e^{-i\Phi_g/2} e^{-i\Phi_h/2} &=& \omega_\Phi(g,h)\,e^{-i\Phi_{gh}/2}\\
        \nonumber\\
        \text{SC state:}\;\;\;\;\;\;\;
        \Tilde{U}^g(h\,\mathbf{k})\:\Tilde{U}^h(\mathbf{k}) &=& \Tilde{\omega}(g,h)\: \Tilde{U}^{gh}(\mathbf{k})
    \end{eqnarray}
\end{subequations}
Eq.~\eqref{U-tilde} relates the three projective representations and eq.~\eqref{wtilde=wphi-w0} relates their cocycles. 

Given the normal state PSG and the pairing symmetry of the SC state, one can use the formalism described above
to determine the SC state PSG. This is achieved in the following steps.  Pairing symmetry being a 1D linear representation, $\mathcal{R}_{\text{pair}}$ can be read off from the character table of $X$. Taking the square roots of the characters one obtains the 1D projective representation $\mathcal{R}_{\Phi}$ and its cocycle $\omega_{\Phi}$. With the normal state PSG known
eq.~\eqref{wtilde=wphi-w0} gives the SC state PSG while eq.~\eqref{U-tilde} gives the SC state projective representation explicitly. 
Thus knowing the pairing symmetry enables us to find the SC state PSG that preserves the BdG Hamiltonian. 
In the next Section we turn to the inverse problem of constraining the pairing symmetry, knowing the SC state PSG.

\section{Constraints on the pairing symmetry by the PSG}\label{sec:PSG-->pairing sym}

One longstanding experimental challenge in the field of superconductivity is how to unambiguously determine the pairing symmetry of a superconductor material, based on experimental measurements. Since all fermionic excitations in the superconductor form a linear representation of the SC state PSG $\tilde X_f$, the low-temperature physical properties of the superconductor completely depend on the PSG. For example, as will be discussed in section \ref{sec:consequence}, the topological properties of the SC phase are determined by the PSG. As a result, it seems plausible to detect the SC state PSG $\tilde X_f$ using various experimental probes, which we will clarify in future publications. This observation motivates us to answer the following question: given a SC state PSG $\tilde X_f$, what are the pairing symmetries compatible with $\tilde X_f$? In other words, how does the a given PSG constrain the possible pairing symmetry in a superconductor? The answer to this question will allow us to constrain or even determine the pairing symmetry of a SC, by experimentally detecting its PSG. 

Based on the discussions in section \ref{sec:pairing sym-->PSG}, we can readily derive the constraints on the pairing symmetry by the PSG from relations (\ref{U-tilde}) and (\ref{wtilde=wphi-w0}). Specifically, given a SC state PSG $\tilde X_f$ and its associated 2-cocycle $\tilde\omega$, we can follow the steps listed below to obtain the possible pairing symmetries $\mathcal{R}_{pair}$ in (\ref{pairing-symm})-(\ref{pairing-symm-2}):

(1) Given the crystalline point group $X$, determine the normal state PSG $X_f^0$ and associated 2-cocycle $\{\omega_0\}$ of the normal-state symmetry transformations $\{U^g_0|g\in X\}$. This only depends on the strength of SOCs in the system. 

(2) Compute the 2-cocycle $\{\omega_\Phi\}$ from $\{\omega_0\}$ and $\{\tilde\omega\}$ from relation (\ref{wtilde=wphi-w0}).

(3) Obtain all one-dimensional (1d) projective representations $\{\mathcal{R}_\Phi(g)|g\in X\}$ of the crystalline symmetry group $X$ compatible with 2-cocyle $\{\omega_\Phi\}$ obtained in step (2), satisfying
\bea
\mathcal{R}_\Phi(g)\mathcal{R}_\Phi(h)=\omega_\Phi(g,h)\mathcal{R}_\Phi(gh)
\eea

(5) For each 1d projective representation $\mathcal{R}_\Phi(g)$ obtained in step (3), compute the 1d linear representation
\bea\label{Rep pair:1d}
\mathcal{R}_{pair}(g)=\big[\mathcal{R}_\Phi(g)\big]^{-2}
\eea
of the pairing order parameter. The collection of all results $\{\mathcal{R}_{pair}\}$ correspond to all the possible pairing symmetries compatible with the PSG $\tilde X_f$.\\

We have applied our general computational scheme to the case of 32 crystalline point groups for both strong SOCs and neglible (weak) SOCs. Group cohomology and projective representation calculations are performed using the GAP computer algebra
program\cite{GAP4}. The correspondence between fermion PSGs $G_f$ and the representations $\mathcal{R}_{pair}$ of the superconducting order parameter is established for all 32 point groups, and the results are summarized in Table.~\ref{table:point_group} in Appendix \ref{app:table}.


\subsection{Examples}\label{sec:examples}
In what follows, we consider three point groups to demonstrate the procedure described above. First we consider $X=C_{4v}$ with both weak and strong SOC. We tabulate the pairing symmetries that are compatible with each PSG of this group. In this case, it turns out that the pairing symmetry can be uniquely identified from the superconducting state PSG. In addition, for each pairing symmetry we produce physical examples of pair wave-functions. This analysis, for instance, is relevant to cuprate like systems. We carry out similar deductions for $X=D_6$ with weak spin orbit coupling which could be important for graphene based systems and $X=C_{3v}$ with strong spin-orbit coupling which relevant for transisition metal dichalcogenides (TMD). 
\subsubsection{Superconductivity on the square lattice with $X=C_{4v}$}

We first demonstrate how to obtain the compatible pairing symmetries from the PSG in an example of point group $C_{4v}$, applicable to superconductivity in the copper-oxygen (Cu-O) plane of layered cuprate superconductors. Denoting the two-dimensional (2d) Cu-O plane as the $x$-$y$ plane, the $C_{4v}$ group is generated by a 90-degree rotation $C_4$ along the $\hat z$ axis and a mirror reflection $\sigma_v$ in the $y$-$z$ plane:
\bea
(x,y,z)\overset{C_4}\longrightarrow(y,-x,z);~~~(x,y,z)\overset{\sigma_v}\longrightarrow(-x,y,z).
\eea

All inequivalent 2-cocycle of $C_{4v}$ group $\{\omega\}\in\mathcal{H}^2(C_{4v},Z_2)$ is classified by the following three $Z_2$-valued gauge-invariant coefficients:
\bea\label{2-cocycle:C4v}
\{\omega\}\equiv\big(
\omega(C_4^2,C_4^2),\omega(\sigma_v,\sigma_v),\omega(\sigma_d,\sigma_d)\big)=(\pm1,\pm1,\pm1)\in\mbz_2^3,
\eea
where $\sigma_d\equiv C_4\sigma_v$.

Alternatively, for the 2-cocycle $\{\tilde\omega\}$ of the superconducting phase, the three coefficients above translate into the following algebraic relations that define the PSG $G_f$:

\bea
(C_4^\prime)^4=\big[\tilde\omega(C_4^2,C_4^2)\big]^{\hat F},~~~(\sigma_v^\prime)^2=\big[\tilde \omega(\sigma_v,\sigma_v)\big]^{\hat F},~~~(\sigma_d^\prime)^2=\big[\tilde\omega(\sigma_d,\sigma_d)\big]^{\hat F}. 
\eea

\begin{table*}[t]
\begin{tabular}{|c||c|c|c|c|c|c|c|c|}
\hline
PSG data $\{\tilde\omega\}$&$(1,1,1)$&$(1,1,-1)$&$(1,-1,1)$&$(1,-1,-1)$&$(-1,1,1)$&$(-1,1,-1)$&$(-1,-1,1)$&$(-1,-1,-1)$\\
\hline
$\{\omega_\Phi\}$&$(-1,-1,-1)$&$(-1,-1,1)$&$(-1,1,-1)$&$(-1,1,1)$&$(1,-1,-1)$&$(1,-1,1)$&$(1,1,-1)$&$(1,1,1)$\\
\hline
1d proj. rep. $\{\mathcal{R}_\Phi\}$&N/A&N/A&N/A&N/A&$(\pm1,\pm\imth)$&$(\pm\imth,\pm\imth)$&$(\pm\imth,\pm1)$&$(\pm1,\pm1)$\\
\hline
$\{\mathcal{R}_{pair}\}$ in (\ref{1d rep:C4v:pair})&N/A&N/A&N/A&N/A&$(1,-1)$&$(-1,-1)$&$(-1,1)$&$(1,1)$\\
\hline
Pairing sym. (w/ SOC)&N/A&N/A&N/A&N/A&$A_{2}$&$B_{2}$&$B_{1}$&$A_{1}$\\
\hline
\end{tabular}
\caption{In the case of $X=C_{4v}$ group with strong SOCs, we summarize the map from the PSG data, i.e. 2-cocycle $\{\tilde\omega\}$ defined in (\ref{2-cocycle:C4v}), to the representation $\{\mathcal{R}_{pair}\}$ of the pairing wave function defined in (\ref{1d rep:C4v:pair}). Here we consider systems with strong SOCs, with $\{\omega_0\}=(-1,-1,-1)$ for the crystalline symmetry representation $\{U^g|g\in X\}$ in the normal state.}\label{tab:C4v:SOC}
\end{table*}

\begin{table*}[t]
\begin{tabular}{|c||c|c|c|c|c|c|c|c|}
\hline
PSG data $\{\tilde\omega\}$&$(1,1,1)$&$(1,1,-1)$&$(1,-1,1)$&$(1,-1,-1)$&$(-1,1,1)$&$(-1,1,-1)$&$(-1,-1,1)$&$(-1,-1,-1)$\\
\hline
$\{\omega_\Phi\}$&$(1,1,1)$&$(1,1,-1)$&$(1,-1,1)$&$(1,-1,-1)$&$(-1,1,1)$&$(-1,1,-1)$&$(-1,-1,1)$&$(-1,-1,-1)$\\
\hline
1d proj. rep. $\{\mathcal{R}_\Phi\}$&$(\pm1,\pm1)$&$(\pm\imth,\pm 1)$&$(\pm\imth,\pm\imth)$&$(\pm1,\pm\imth)$&N/A&N/A&N/A&N/A\\
\hline
$\{\mathcal{R}_{pair}\}$ in (\ref{1d rep:C4v:pair})&$(1,1)$&$(-1,1)$&$(-1,-1)$&$(1,-1)$&N/A&N/A&N/A&N/A\\
\hline
Pairing sym. (no SOC)&$A_{1}$&$B_{1}$&$B_{2}$&$A_{2}$&N/A&N/A&N/A&N/A\\
\hline
\end{tabular}
\caption{In the case of $X=C_{4v}$ group with negligible SOCs, we summarize the map from the PSG data, i.e. 2-cocycle $\{\tilde\omega\}$ defined in (\ref{2-cocycle:C4v}), to the representation $\{\mathcal{R}_{pair}\}$ of the singlet pairing wave function defined in (\ref{1d rep:C4v:pair}). Here we consider systems with negligibly weak SOCs and approximate $SU(2)$ spin rotational symmetries, with $\{\omega_0\}=(1,1,1)$ for the crystalline symmetry representation $\{U^g|g\in X\}$ in the normal state.}\label{tab:C4v:no SOC}
\end{table*}

The one-dimensional (1d) projective representation of the $X=C_{4v}$ group is generally given by 
\bea\label{1d rep:C4v:Phi}
\{\mathcal{R}_\Phi\}\equiv(\mathcal{R}_\Phi({C_4})\in U(1),\mathcal{R}_\Phi({\sigma_v})\in U(1))
\eea
When we consider all 8 iequivalent 2-cocycles (\ref{2-cocycle:C4v}) for $C_{4v}$ group, only 4 of them admit 1d projective representations, as shown in Table \ref{tab:C4v:SOC}. This is due to the following property of 1d representations: 
\bea
[\mathcal{R}_\Phi(C_4)]^4=\omega_\Phi(C_4^2,C_4^2)=[\mathcal{R}_\Phi({\sigma_d})]^4/[\mathcal{R}_\Phi(\sigma_v)]^4=\big[\omega_\Phi(\sigma_d,\sigma_d)/\omega_\Phi(\sigma_v,\sigma_v)\big]^2\equiv1.
\eea
Therefore, only four of the eight 2-cocycles with $\omega_\Phi(C_4^2,C_4^2)=1$ are compatible with 1d projective representations, while the projective representations associated with the other four 2-cocoyles with $\omega_\Phi(C_4^2,C_4^2)=-1$ are at least two-dimensional. Next, making use of the homomorphism $\rho$ defined in (\ref{Rep pair:1d}), we can obtain the group representation
\bea\label{1d rep:C4v:pair}
\{\mathcal{R}_{pair}\}\equiv(\mathcal{R}_{pair}(C_4),\mathcal{R}_{pair}(\sigma_v)
)
\eea
of the pairing wave function, and determine the pairing symmetry compatible with the PSG $G_f$. We summarize the results in Table \ref{tab:C4v:SOC} for superconductors with strong SOCs, and in Table \ref{tab:C4v:no SOC} for singlet superconductors with negligible SOCs. 

To provide illustrative examples, consider the simple example of spin-$\frac{1}{2}$ fermions on a 2D square lattice with a single orbital at each site. In eqn. \eqref{ham:pairing} the indices $\alpha,\beta$ now run over spin labels $\uparrow,\downarrow$. Then the order parameter takes the form,
\begin{eqnarray}
       \Delta(\mathbf{k}) =\begin{pmatrix}
        \Delta_{\uparrow\uparrow}(\mathbf{k})&&\Delta_{\uparrow\downarrow}(\mathbf{k})\\ \Delta_{\downarrow\uparrow}(\mathbf{k})&&\Delta_{\downarrow\downarrow}(\mathbf{k})
    \end{pmatrix}
    \label{Delta2}
\end{eqnarray} 
Pauli exclusion enforces the constraint $\Delta_{ss'}(\mathbf{k}) = -\Delta_{s's}(-\mathbf{k})$ which implies that the diagonal entries are odd functions of $\mathbf{k}$ while the off-diagonal entries are related by $\Delta_{\downarrow\uparrow}(\mathbf{k}) = - \Delta_{\uparrow\downarrow}(-\mathbf{k})$. Spin rotation leaves the spatial degrees of freedom unaffected and transforms the fermion operators as $S_\theta c_{\mathbf{k},s} S^{-1}_\theta = \left[ e^{-i\frac{\theta}{2}\hat{n}.\Vec{\sigma}} \right]_{ss'}c_{\mathbf{k},s'}$. In the weak spin orbit coupling regime, the order parameter is spin singlet and the pair wave function takes the form
\begin{equation}
    \Delta(\mathbf{k}) = \psi(\mathbf{k})(i\sigma_y)
    \label{weak-soc}
\end{equation}
where $\psi(-\mathbf{k}) = \psi(\mathbf{k})$ by Pauli exclusion. In the strong spin orbit coupling limit however, when spin rotation invariance is lost, pairing could have a superposition of singlet and triplet terms and \eqref{Delta2} is often recast in the following form showing the two parts explicitly.
\begin{equation}
    \Delta(\mathbf{k}) = \psi(\mathbf{k})(i\sigma_y) + \Vec{d}(\mathbf{k})\cdot\Vec{\sigma}(i\sigma_y)
    \label{stron-soc}
\end{equation}
Where $\psi(\mathbf{k}) = \frac{1}{2}(\Delta_{\uparrow\downarrow}(\mathbf{k})+ \Delta_{\uparrow\downarrow}(-\mathbf{k}))$, $d_z(\mathbf{k}) =  \frac{1}{2}(\Delta_{\uparrow\downarrow}(\mathbf{k})- \Delta_{\uparrow\downarrow}(-\mathbf{k}))$, $d_x(\mathbf{k}) = \frac{1}{2}(\Delta_{\downarrow\downarrow}(\mathbf{k})- \Delta_{\uparrow\uparrow}(-\mathbf{k}))$, $d_y(\mathbf{k}) = \frac{1}{2i}(\Delta_{\downarrow\downarrow}(\mathbf{k})+ \Delta_{\uparrow\uparrow}(-\mathbf{k}))$. Evidently, the constraint imposed by Pauli exclusion now translates to $\psi(-\mathbf{k})=\psi(\mathbf{k})$ and $\Vec{d}(-\mathbf{k})=-\Vec{d}(\mathbf{k})$.

Assuming that the unbroken symmetry group in the superconducting state is $X =C_{4v}$, we first discuss the normal state projective representation $U_0^g(\mathbf{k})$ and the corresponding PSG. The normal state PSG is different in the weak and strong SOC limit. In each of these two limits we produce examples of $\Delta^I(\mathbf{k})$ such that the pairing term transforms as the I-th one-dimensional irreducible representation of $X$ under the normal state symmetry transformations \cite{sigrist1991}.
\begin{equation}
U_0^g(\mathbf{k})\Delta^I(\mathbf{k})\left[U_0^g(-\mathbf{k})\right]^T = e^{i\Phi^I_g} \Delta^I(g\mathbf{k})
\label{pairingsym}
\end{equation}
 As discussed earlier $\left\{e^{i\Phi^I_g}\right\}$ is precisely the I-th 1D linear irrep of X which we call pairing symmetry $\mathcal{R}_{pair}$ and it can be read off from the character table. Note that for the group $C_{4v}$, any 1D linear irrep is completely determined by the characters for $C_4$ rotation and vertical mirror $M_x$, $(e^{i\Phi_{C_4}},e^{i\Phi_{M_x}})$. 
 

In the weak SOC limit, spatial symmetries do not affect spin labels, i.e, $\hat{g} \hat{c}_{\mathbf{k},s} \hat{g}^{-1} = \hat{c}_{g\mathbf{k},s}$. As a result, in this case $U_0^g(\mathbf{k}) = 1$ and the projective representation is characterized by the trivial PSG $\omega_0 \equiv (1,1,1)$. Examples of $\psi(\mathbf{k})$ for the different pairing symmetries are an $s$-wave for $A_1$, a $g$-wave for $A_2$, a $d_{x^2-y^2}$-wave for $B_1$ and a $d_{xy}$-wave for $B_2$. 

In the strong SOC limit, the symmetry operations act both on the spatial and spin degrees of freedom and the normal state projective representation is $U^{C_4}(\mathbf{k}) = e^{i\frac{\pi}{4}\sigma_z} $ and $U^{M_x}(\mathbf{k}) = i\sigma_x$. The latter is because the vertical mirror $M_x$ can be thought of as a product of inversion $I$ and $C_{2x}$ i.e, a rotation by $\pi$ around the x-axis. The spinor representations of $I$ and $C_{2x}$ can then be chosen to be $1$ and $e^{i\frac{\pi}{2}\sigma_x} = i\sigma_x$ respectively. As a consequence of this, the representation for $C_{2z}$ is $U^{C_2}(\mathbf{k}) = e^{i\frac{\pi}{2}\sigma_z} = i\sigma_z$. Also, note that while we are in two dimensions, $C_2 \mathbf{k} = -\mathbf{k}$ and the character for $C_2$ in all the irreps is $+1$. Plugging all of this information in eqn \eqref{pairingsym} for $g=C_2$ then tells us that $\Delta_{\uparrow\downarrow}(\mathbf{k})$ is even in $\mathbf{k}$ and hence $d_z(\mathbf{k}) = 0$. In the examples that we construct, we set $\psi(\mathbf{k}) = 0$ so that pairing is purely triplet. Thus $\Delta^I(\mathbf{k}) = i (d^I_y(\mathbf{k}) \mathds{1} + i d^I_x(\mathbf{k})\sigma_z)$ and in terms of the $d$-vector, eqn. \eqref{pairingsym} reduces to $(d^I_y(\mathbf{k}),-d^I_x(\mathbf{k})) = e^{i\Phi^I_{C_4}} (d^I_x(C_4\mathbf{k}),d^I_y(C_4\mathbf{k}))$ and $(d^I_x(\mathbf{k}),-d^I_y(\mathbf{k})) = e^{i\Phi^I_{M_x}} (d^I_x(M_x\mathbf{k}),d^I_y(M_x\mathbf{k}))$. Examples in this category are of type $(p+ip)\uparrow + (p-ip)\downarrow$ and the respective $d$-vectors are shown in table .
\begin{table}[]
\centering
    \begin{tabular}{||c||c|c|c|c||}
    \hline\hline
    $\mathcal{R}_{pair} = (e^{i\Phi_{C_4}},e^{i\Phi_{M_x}}) $ & $\psi(\mathbf{k})$& $\Vec{d}(\mathbf{k})$ & $\mathcal{R}_{\Phi} = (e^{-i\Phi_{C_4}/2},e^{-i\Phi_{M_x}/2})$&$\omega_\Phi$ \\
    \hline\hline
         $A_{1} = (+1,+1)$& $1$ & $(-k_y,+k_x,0)$ &$(\pm 1,\pm 1)$ & $(+1,+1,+1)$ \\
         \hline
         $A_{2} = (+1,-1)$& $k_xk_y(k_x^2-k_y^2)$ &$(+k_x,+k_y,0)$&$(\pm1,\pm i)$& $(+1,-1,-1)$  \\
         \hline
         $B_{1} = (-1,+1)$& $(k_x^2 - k_y^2)$ & $(+k_y,+k_x,0)$&$(\pm i, \pm 1)$ & $(+1,+1,-1)$\\
         \hline
         $B_{2} = (-1,-1)$& $k_x k_y$& $(+k_x,-k_y,0)$& $(\pm i, \pm i)$ & $(+1,-1,+1)$ \\
     \hline    
    \end{tabular}
    \caption{singlet and triplet pair wave-functions corresponding to the 1D irreps of $C_4v$. The 1D projective representation $\mathcal{R}_\Phi$ and its corresponding cocycle $\omega_\Phi$ are also tabulated.(see main text for details)}
\label{table:exam-C4v}
\end{table}
\subsubsection{Singlet superconductivity on the honeycomb lattice with $X=D_{6}$}
Our next example is the point group $D_6$ applicable to superconductivity in the magic-angle twisted bilayer graphene. Let's denote the 2d lattice plane as the $x$-$y$ plane. The point group $D_6$ is generated by a $C_2$ rotation along $\hat{x}$ axis and a $C_6$ rotation along $\hat{z}$ axis:
\bea
(x,y,z)\overset{C_6}\longrightarrow(\frac{1}{2}x-\frac{\sqrt{3}}{2}y,\frac{\sqrt{3}}{2}x+\frac{1}{2}y,z);~~~(x,y,z)\overset{C_{2x}}\longrightarrow(x,-y,-z).
\eea

The cohomology classification is $\mathcal{H}^2(D_6,Z_2)=\mathbb{Z}_2^3$ and can be characterized by the following three invariant cocycles: 
\bea
\{\omega  \}\equiv (\omega(C_{2z},C_{2z}),\omega(C_{2x},C_{2x}),\frac{\omega(C_{2z},C_{2x})}{\omega(C_{2x},C_{2z})}),
\eea
where the last factor must be equal to $+1$ in order to admit 1d projective representations due to the fact that:
\bea
\omega(C_{2z},C_{2x})\mathcal{R}_{\Phi}(C_{2y})=\mathcal{R}_{\Phi}(C_{2z})\mathcal{R}_{\Phi}(C_{2x})=\mathcal{R}_{\Phi}(C_{2x})\mathcal{R}_{\Phi}(C_{2z})=\omega(C_{2x},C_{2z})\mathcal{R}_{\Phi}(C_{2y}).
\eea

As we are considering singlet superconductor with negligible SOCs in which case we have $\{\omega_0\}=(1,1,1).$ 

When we restrict ourselves to two spatial dimensions as is the case here, rotations by $\pi$ about an in plane axis, for example $C_{2y}$, can be regarded as reflection about the mirror $M_x$ perpendicular to the plane and hence the action groups $D_6$ and $C_{6v}$ cannot be distinguished on the 2D plane.  

The group has four one dimensional irreducible representations, each of which has a $C_3$ character of $+1$. Moreove, the four 1D irreps are uniquely identified by the $C_{2z}$ and $C_{2x}$ characters, $(e^{i\Phi_{C_{2z}}},e^{i\Phi_{C_{2x}}}) = (\pm1,\pm1)$. Like in the previous example, the four pairing symmetries are in a one to one correspondence with the four PSG s $\omega_\Phi = (\pm 1,\pm 1,1)$ of the 1D projective representations $\mathcal{R}_{\Phi}$. This has been illustrated in table.

To give illustrative examples, let us start with the hypothetical situation  of singlet superconductivity in mono-layer graphene, assuming that the ground state inherits the $D_6$ crystalline symmetry of the normal state. The $\alpha$, $\beta$ indices in eqn. \eqref{ham:pairing} now run over spin ($\uparrow,$$\downarrow$) and sub-lattice labels ($A,B$). Assuming that it is a spin singlet superconductor, the pairing term has the general form,
\begin{eqnarray}
        \Delta(\mathbf{k}) &=& \begin{pmatrix}
        \Delta_{AA}(\mathbf{k})&&\Delta_{AB}(\mathbf{k})\\ \Delta_{BA}(\mathbf{k})&&\Delta_{BB}(\mathbf{k}) 
    \end{pmatrix}(is_y)
    \label{Delta-MLG}
\end{eqnarray}
Where the Pauli matrix $s_y$ acts on the spin degrees of freedom. Fermi statistics demands $\Delta^T(-\mathbf{k}) = -\Delta(\mathbf{k})$ and as a consequence $\Delta_{AA}(\mathbf{k})$, $\Delta_{BB}(\mathbf{k})$ are even functions of $\mathbf{k}$ and $\Delta_{BA}(\mathbf{k}) = \Delta_{AB}(-\mathbf{k})$. The projective representation of $D_6$ that preserves the normal state hamiltonian is 
\begin{equation}
    U_0^{C_3}(\mathbf{k}) = 1;  U_0^{C_{2z}}(\mathbf{k}) = \sigma_x; U_0^{C_{2x}}(\mathbf{k}) = 1
    \label{norm-PSG}
\end{equation}
Where Pauli matrix $\sigma_x$ acts on the sublattice space. It follows that the corresponding PSG is trivial one, $\omega_0 = (1,1,1)$. Like before, we construct examples of $\Delta^I(\mathbf{k})$ satisfying eqn. \eqref{pairingsym} for each pairing symmetry $\mathcal{R}_{pair} \equiv (e^{i\Phi_{C_{2z}}},e^{i\Phi_{C_{2x}}})$. The 1D projective representation $\mathcal{R}_{\Phi}=(e^{-i\Phi_{C_{2z}}/2},e^{-i\Phi_{C_{2x}}/2})$ and its PSG $\omega_\Phi$ are also obtained to show the one to one correspondence as noted earlier.

We end this discussion with the case of superconductivity in magic angle twisted bilayer graphene (MATBG). It was argued in \cite{lake2022} that superconductivity in MATBG descends from a spin valley locked normal state which breaks spin rotation invariance spontaneously. While the normal state has $D_6$ symmetry, it is spontaneously broken in the superconducting phase as indicated by the observation of nematicity. Thus condensation must take place in either the $E_1$ or the $E_2$ irrep of $D_6$. Correspondingly the orbital part of the pair wave function is a $p$-wave or a $d$-wave respectively. Since the $p$-wave has a $C_2$ character $-1$ and the $d$-wave has a $C_2$ character $+1$, these correspond to different $\mathcal{R}_\text{pair}$ s of the residual symmetry group of the superconductor $X=C_{2z}$. There is a one to one correspondence between $\mathcal{R}_\text{pair}$ and PSG for $X$ as shown in table in appendix \ref{app:table} . Thus experimental detection of a physical manifestation of the PSG that can distinguish between these two cases will allow us to say if the orbital part is a $p$-wave or a $d$-wave.

\begin{table}[t]
    \begin{tabular}{||c||c|c|c|c|c||}
    \hline\hline
    $\mathcal{R}_{pair} = (e^{i\Phi_{C_{2z}}},e^{i\Phi_{C_{2x}}}) $ & $\Delta_{AA}(\mathbf{k})$& $\Delta_{BB}(\mathbf{k})$ & $\Delta_{AB}(\mathbf{k})$ &$\mathcal{R}_{\Phi} = (e^{-i\Phi_{C_{2z}}/2},e^{-i\Phi_{C_{2x}}/2})$&$\omega_\Phi$ \\
    \hline\hline
         $A_{1} = (+1,+1)$& $\Delta_0$ & $\Delta_0$ & $\Delta_0'$ &$(\pm 1,\pm 1)$ & $(+1,+1,+1)$ \\
        \hline
         $A_{2} = (+1,-1)$& $\Delta_0 f(k_x,k_y)$ & $\Delta_0 f(k_x,k_y)$& $\Delta'_0 g(k_x,k_y)$&$(\pm1,\pm i)$& $(+1,-1,+1)$  \\
         \hline
         $B_{1} = (-1,+1)$& $\Delta_0$ & $-\Delta_0$&$0$&$(\pm i, \pm 1)$ & $(-1,+1,+1)$\\
         \hline
         $B_{2} = (-1,-1)$& $\Delta_0 f(k_x,k_y)$& $-\Delta_0 f(k_x,k_y)$&$0$& $(\pm i, \pm i)$ & $(-1,-1,+1)$ \\
     \hline

    \end{tabular}
    \caption{ Examples of pair wave function with different possible pairing symmetries on a honeycomb lattice which has point group $D_6$. The 1D projective representations $\mathcal{R}_\Phi$ and corresponding cocycles $\omega_\Phi$ are also listed. We have used the notation $f(k_x,k_y) = k_xk_y(k_x^2-3k_y^2)(k_y^2-3k_x^2)$ and $g(k_x,k_y) = k_y(3k_x^2-k_y^2)$}
\end{table}
\subsubsection{Superconductivity in transition metal diacalcogenide (TMD) with $X=C_{3v}$}
Our last example is the point group $C_{3v}$ applicable to the case of superconductivity in monolayer transition metal diacalcogenide. Let's denote the 2d plane as the $x$-$y$ plane, the $C_{3v}$ group is generated by a 3-fold rotation $C_3$ along $\hat{z}$ axis and a mirror reflection $M_x$ in the $y$-$z$ plane:
\bea
(x,y,z)\overset{C_3}\longrightarrow(-\frac{1}{2}x-\frac{\sqrt{3}}{2}y,\frac{\sqrt{3}}{2}x-\frac{1}{2}y,z);~~~(x,y,z)\overset{M_x}\longrightarrow(-x,y,z).
\eea

The cohomology classification $\mathcal{H}^2(C_{3v},Z_2)$ is particularly simple: it is $\mathbb{Z}_2$ and characterized by ${\omega}(M_x,M_x)=\pm1$. As monolayer TMDs has strong SOCs, we have $\omega_0(M_x,M_x)=-1$.

In monolayer TMD s, strong Ising SOC along with in-plane inversion symmetry breaking gives rise to a band structure where the degeneracy between the z-polarized spin states, $\ket{\uparrow}$ and $\ket{\downarrow}$ is lifted \cite{xi2016}. Moreover, due to time reversal symmetry, the energies of the $\ket{\uparrow}$ and $\ket{\downarrow}$ bands are reverse in the two valleys. At small doping, the Fermi arcs are two concentric rings at each valley with opposite spin polarization. Added to that, since the splitting between the spin states is opposite in the two valleys, spin and valley are completely correlated. In other words, if the inner Fermi arc at K valley has spin polarization $\ket{\uparrow}$, the inner Fermi arc at the K' valley has spin polarization $\ket{\downarrow}$. The outer Fermi arcs then have polarizations $\ket{\downarrow}$ and $\ket{\uparrow}$ at K and K' valleys respectively. For cooper pairs with zero center of mass momentum, pairing must occur between inner or outer Fermi arcs of opposite valleys. Owing to "spin-valley locking", pairs between the inner (or outer) Fermi arcs say can only involve $\ket{\uparrow}$ electrons from K-valley and $\ket{\downarrow}$ electrons from the $K'$ valley. This severely restricts the form of pairing. 

We then write the general $4\times 4$ pairing matrix $\Delta(\mathbf{k})$ written in spin ($s$) and valley ($\tau$) degrees of freedom, consistent with Fermi statistics, $\Delta^T(-\mathbf{k}) = -\Delta(\mathbf{k})$ and spin-valley locking as enforced by $\mathcal{P}^T \Delta(\mathbf{k}) = \Delta(\mathbf{k}) \mathcal{P} = \Delta(\mathbf{k}) $, where $\mathcal{P} = \frac{1}{2}(1+\tau_zs_z)$ projects onto the $\tau_z s_z = 1$ subspace.
\begin{equation}
    \Delta(\mathbf{k}) = \left[ \left(f_{\mathbf{k}}\tau_+ - f_{-\mathbf{k}}\tau_-\right)\hat{z}.\Vec{s} + \left(f_{\mathbf{k}}\tau_+ + f_{-\mathbf{k}}\tau_-\right)\right](is_y)
    \label{tmd_delta}
\end{equation}

This shows that pairing is a superposition of singlet and triplet ($S_z = 0$) channels and is characterized by a single function $f_{\mathbf{k}}$. The projective representation of $C_{3v}$ that preserves the normal state hamiltonian is 
\begin{equation}
    U_0^{M_x}(\mathbf{k}) = is_x\tau_x ; U_0^{C_3}(\mathbf{k}) = e^{i\frac{\pi}{3} s_z} 
\end{equation}
Evidently, this corresponds to $\omega_0(M_x,M_x)=-1$ Plugging \eqref{tmd_delta} into \eqref{pairingsym}, the condition for $\Delta(\mathbf{k})$ to transform as a one-dimensional irrep of $C_{3v}$ translates to $f(\mathbf{k}) = f(C_3\mathbf{k})$ and $f(-\mathbf{k}) = e^{i\Phi_{M_x}}f(M_x\mathbf{k})$. 

\begin{table}[h]
    \begin{tabular}{||c||c|c|c||}
    \hline\hline
    $\mathcal{R}_{pair} = e^{i\Phi_{M_{x}}} $& $f_{\mathbf{k}}$ &$\mathcal{R}_{\Phi} = e^{-i\Phi_{M_{x}}/2}$&$\omega_\Phi$ \\
    \hline\hline
         $A_{1} = +1$& $\Delta_0$ &$\pm 1$ & $+1$ \\
        \hline
         $A_{2} = -1$& $\Delta_0k_y(3k_x^2-k_y^2)$ &$\pm i$& $-1$  \\
         
     \hline

    \end{tabular}
    \caption{Examples of pair wave function corresponding to the two 1D irreps of $C_{3v}$ which are relevant to superconductivity in transition metal dichalcogenides (TMDs) as discussed in the text. The 1D projective representation and corresponding cocycles are also listed}
\end{table}

\subsection{Physical consequences of the PSG}\label{sec:consequence}

The projective symmetry group $G_f$ of the BdG Hamiltonian has effects on all fermionic excitations of the superconductor, since the Bogoliubov quasipaticles as excitations of the BdG Hamiltonian form a linear representation of the PSG $G_f$. In particular, the topological properties of the superconductor is determined by the PSG, as different PSGs give rise to different classifications of fermion topological superconductors (TSCs)\cite{chiu2016,Wang2020,shiozaki2022classification}. This is a well-known fact in the classification of gapped fermion topological phases, both in the 10-fold way\cite{Altland1997} classification of non-interacting topological superconductors\cite{Schnyder2008,chiu2016}, and in the interacting classification of fermion symmetry protected topological phases\cite{Kapustin2015,Wang2020}. For example, in the case of time reversal symmetry $\bst$, it is well known that two- and three-dimensional topological insulators only exist for spinful electrons with $\bst ^2=(-1)^{\hat F}$ and $G_f=U(1)\rtimes Z_4^\bst$, which is a different symmetry class (class AII in the 10-fold way\cite{Altland1997}) than spinless case (class AI in the 10-fold way\cite{Altland1997}), with $\bst^2=1$ and $G_f=U(1)\rtimes Z_2^\bst$. In addition to topological classifications, these two distinct symmetry classes have many other different properties, such as the presence vs. absence of Kramers degeneracy of fermion excitations. Below we illustrate how different PSGs, and hence different pairing symmetries, give rise to different classifications of TSCs, in the case of crystalline symmetries\cite{Ando2015,Okuma2019,shiozaki2022classification,Ono2022,Zhang2022,Shiozaki2023,Ono2023}. We use 3d SCs with mirror reflection symmetry $M_x$, and 2d SCs with 2-fold rotational symmetry $C_{2z}$ as two known examples to demonstrate this fact. 

\subsubsection{3d SCs with mirror reflection symmetry $M_x$}

Our first example is the classification of TSCs in three dimension (3d) in the presence of only mirror reflection symmetry $M_x$ which reverses the $x$ coordinate. From the group cohomology $\mathcal{H}^2(M_x,Z_2)=Z_2$, we find two possible fermion PSGs in the presence of strong SOCs: $G_f=Z_2^{\tilde M_x}\times Z_2^F$ with $\tilde M_x^2=+1$, and $G_f=Z_4^{\tilde M_x}$ with $\tilde M_x^2=(-1)^{\hat F}$. Similarly, in the presence of weak SOCs and $SU(2)$ spin rotational symmetry, the two possible PSGs are given by $G_f=SU(2)\times Z_2^{\tilde M_x}$ with $\tilde M_x^2=+1$, and $G_f=SU(2)\times Z_4^{\tilde M_x}/Z_2$ with $\tilde M_x^2=(-1)^{\hat F}$. 

For weakly interacting systems, $K$-theory\cite{kitaev2009periodic,Schnyder2008,chiu2013mirror,morimoto2013mirror,chiu2016} can be used to classify distinct TSCs described by BdG Hamiltonians. In the presence of strong SOCs, it gives rise to a $\mathbb{Z}$ classification of TSCs for the case of $\tilde M_x^2=+1$, and a trivial classification for the case of $\tilde M_x^2=(-1)^{\hat F}$\cite{morimoto2013mirror,chiu2013mirror}. In the presence of a weak SOC and $SU(2)$ spin rotational symmetry, there is a $\mathbb{Z}$ classification of TSCs for the case of $\tilde M_x^2=+1$, and a $\mbz_2$ classification for the case of $\tilde M_x^2=(-1)^{\hat F}$\cite{morimoto2013mirror,chiu2013mirror}. With this result we can now readily bridge the gap between pairing symmetry and the K-theory classification of TSC via the projective symmetry group $G_f$. 

A mirror symmetry satisfying $\tilde M_x^2=\mathbbm{1}$ is preserved either in a singlet superconductor with pairing symmetry $A'$ in the presence of a weak SOC, or in a superconductor with pairing symmetry $A''$ in the presence of a strong SOC. The classifications of weakly-interacting TSCs in these two cases  are both $\mathbb{Z}$.

To compare, a mirror symmetry satisfying $\tilde M_x^2=(-1)^{\hat F}$ corresponds to either a singlet superconductor with pairing symmetry $A''$ in the presence of a weak SOC, or a pairing symmetry $A'$ in the presence of a strong SOC. For these two cases the classifications of TSCs are $\mbz_2$ and trivial, respectively. The results are summarized in Table~\ref{table:example-classification-Mx}.

\begin{table}
\begin{tabular}{|c|c|c|c|}
    \hline
       SOC strength &    pairing symmetry& $G_f$   &K-theory classification\cite{chiu2013mirror,morimoto2013mirror}\\
       \hline

   \multirow{2}{*}{Weak}
 & 
$A'$& $SU(2)\times Z_2^{\tilde M_x}$&$\mathbb{Z}$\\ 
\cline{2-4}
 & 
$A''$& $SU(2)\times Z_4^{\tilde M_x}/Z_2$&$\mbz_2$\\ 
\hline
\multirow{2}{*}{Strong}&$A'$&$Z_4^{\tilde M_x}$&0\\
\cline{2-4}
            &$A''$&$Z_2^{\tilde M_x}\times Z_2^F$&$\mathbb{Z}$\\ 

            \hline

\end{tabular}
\caption{Classification of class D topological superconductor in 3d with mirror reflection $M_x$. The fermion projective symmetry groups $G_f$ are listed for superconductors with weak/strong SOCs and $A'$/$A''$ pairing symmetries. Note that the topological classification is solely determined by $G_f$.}\label{table:example-classification-Mx}
       \end{table}

       \begin{table}
\begin{tabular}{|c|c|c|c|}
    \hline
       SOC strength &    pairing symmetry& $G_f$   &K-theory classification\cite{shiozaki2014c2z,Lu2014} \\
       \hline

 \multirow{2}{*}{Weak}
 & 
$A$& $SU(2)\times Z_2^{\tilde C_{2z}}$&$\mathbb{Z}$\\ 
\cline{2-4}
 & 
$B$& $SU(2)\times Z_4^{\tilde C_{2z}}/Z_
2$&$\mathbb{Z}^2$\\ 
\hline
\multirow{2}{*}{Strong}&$A$&$Z_4^{\tilde C_{2z}}$&$\mathbb{Z}^2$\\
\cline{2-4}
            &$B$&$Z_2^{\tilde C_{2z}}\times Z_2^F$&$\mathbb{Z}$\\ 
       \hline     
\end{tabular}
\caption{Classification of class D topological superconductor in 2d with $C_{2z}$ rotation perpendicular to the 2d $x$-$y$ plane. The fermion PSGs $G_f$ are listed for superconductors with weak/strong SOC and $A$/$B$ pairing symmetries. Note that the topological classification is solely determined by $G_f$.}\label{table:example-classification-C2z}
       \end{table}

\subsubsection{2d SCs with 2-fold rotational symmetry $C_{2z}$}

Our second example is the classification of TSCs in two dimensions (2d) with a $C_{2z}$ rotation perpendicular to the 2d plane. In this case $\mathcal{H}^2(C_{2z},Z_2)=Z_2$, which yields two different fermion PSGs in the presence of a strong (weak) SOC: one with $\tilde C_{2z}^2=+1$ and the other with $\tilde C_{2z}^2=(-1)^{\hat F}$, as shown in Table \ref{table:example-classification-C2z}. Accordingly, the $K$-theory classification of $C_{2z}$ symmetric TSCs\cite{shiozaki2014c2z} are given by $\mathbb{Z}$ for $\tilde C_{2z}^2=+1$ and $\mathbb{Z}^2$ for $\tilde C_{2z}^2=(-1)^{\hat F}$.

From the relationship between pairing symmetry and projective symmetry group, we find that the $\tilde C_{2z}^2=+1$ case corresponds to either a singlet SC with pairing symmetry $A$ or a SC with a strong SOC and pairing symmetry $B$. Then for these two cases the classifications of topological superconductors are both $\mathbb{Z}$.

The $\tilde C_{2z}^2=(-1)^{\hat F}$ case corresponds to either a singlet superconductor with pairing symmetry $B$ or a superconductor with a strong SOC and pairing symmetry $A$. For these two cases the classifications of TSCs are both $\mathbb{Z}^2$. The results are summarized in Table~\ref{table:example-classification-C2z}.

From these two examples, we see that BdG Hamiltonians with different PSGs generally give rise to different topological classifications. Based on the correspondence between the fermion PSG and the pairing symmetry discussed in sections \ref{sec:psg}-\ref{sec:PSG-->pairing sym}, the classification of TSCs is therefore directly related to the pairing symmetry, as demonstrated in Table \ref{table:example-classification-Mx}-\ref{table:example-classification-C2z}. For TSCs of all the possible pairing symmetries associated with a magnetic point group symmetry, Ref.\cite{shiozaki2022classification} summarizes a full list of K-theory classification for both the cases of spinless (weak SOC) and spinful (strong SOC) electrons.

\section{General framework}\label{sec:general}

In previous sections, we discussed the systems where the normal state and the superconducting state share the same spin rotational symmetry. Specifically, we focused on two classes of systems: (1) weakly spin orbit coupled which possess spin rotation invariance and (2) Strongly spin orbit coupled cases without any spin rotation invariance. In either case, the fermion PSGs are captured by the central extensions of the point group $X$ (\ref{short exact sequence}) and there is a relatively simple relation between the PSG and the pairing symmetry as discussed in Section \ref{sec:pairing sym-->PSG}.

In the most general situation, the normal state and the superconducting state may have different global (onsite) spin-rotational symmetries, due to spontaneous breaking of spin rotational symmetries by superconductivity. One famous example of such is spin-triplet superconductivity in $^3$He\cite{Volovik2003B}. 
To incorporate these cases, below we describe a generic theory framework that can be applied to all superconductors. 

\subsection{Group extension and pairing symmetry in a generic superconductor}\label{sec:general extension}

Generally, the normal state preserves a global (onsite) spin rotational symmetry group $S_0\subseteq SO(3)$, which is spontaneously broken down to a subgroup $S\subseteq S_0$. Meanwhile, the charge $U(1)$ symmetry in the normal state is completely broken in the superconducting state, leads to a $G=X\times S$ symmetry of the superconductor, where $X$ denotes the spatial/crystalline symmetry group preserved by the superconducting phase. However, as we have already seen in previous discussions, in a fermion system described by BdG Hamiltonian (\ref{ham:BdG}), the full fermion symmetry group $G_f$ of the BdG Hamiltonian is different from the ``bosonic symmetry group'' $G=X\times S$ that act on local operators with an even fermion parity. To describe the fermion symmetry group $G_f$, we have to introduce the mathematical concept of group extension, which is briefly reviewed in Appendix \ref{app:psg}. 

First of all, the fermion global symmetry group $S_f$ is given by the following central extension of spin rotational symmetry $S$:
\bea\label{central extension:onsite sym}
1\rightarrow Z_2^F\rightarrow S_f\rightarrow S\rightarrow 1
\eea
where the center $Z_2^F$ represents the order-2 fermion parity group. Note that the fermion parity group $Z_2^F$ is a normal subgroup of $S_f$. Physically, each spin rotation symmetry $s\in S$ of fermions in the BdG Hamiltonian is a combination of a normal-state spin rotation $U^s$ represented by a $SU(2)$ matrix, and a $U(1)$ charge rotation $e^{\imth\phi_s}$:
\bea
\tilde{U}^s=e^{\imth\phi_s}U^s,~~~U^s\in SU(2),~~~\forall~s\in S.
\eea
such that
\bea
\tilde{U}^s\hat H_\text{BdG}(\tilde{U}^s)^\dagger=\hat H_\text{BdG},~~~\forall~s\in S.
\eea
The $U(1)$ phase factors $\{e^{\imth\phi_s}|s\in S\}$ satisfy
\bea
e^{\imth(\phi_s+\phi_{s^\prime})}=\omega(s,s^\prime)e^{\imth\phi_{ss^\prime}},~~~\omega(s,s^\prime)=\pm1\in Z_2,~~~\forall~s,s^\prime\in S.
\eea
A generic element of fermion onsite symmetry group $S_f$ can be denoted as
\bea
(s,\eta)=\eta^{\hat F}e^{\imth\phi_s\hat F}\hat s,~~~\hat s\in S,~\eta=\pm1.
\eea
where $\hat F$ denotes the total fermion number. Here $\{\tilde U^s|s\in S\}$ form a projective representation of group $S$, or equivalently $\{\tilde U^{(s,\eta)}=\eta\tilde U^s|s\in S,\eta=\pm1\}$ form a linear representation of group $S_f$. Mathematically, different fermion onsite symmetry groups $S_f$, i.e. different central extensions (\ref{central extension:onsite sym}), are classified by 2nd cohomology group $\mathcal{H}^2(S,Z_2^F)$, as discussed in Appendix \ref{app:psg}. 

Next we address the full fermion symmetry group $G_f$ by taking into account of crystalline space group $X$. The fermion symmetry group $G_f$ is given by the extension of space group $X$ by the onsite fermion symmetry group $S_f$:
\bea\label{central extension:space group}
1\rightarrow S_f\rightarrow G_f\rightarrow X\rightarrow 1
\eea
satisfying $X=G_f/S_f$. Similar to previous discussions, we consider the crystalline symmetry group $X$ which is preserved by the superconducting state (hence also by the normal state). Under the crystalline symmetry operation $g\in X$, the electrons in the normal state transform as
\bea
\hat gc_{k,\alpha}\hat g^{-1}=\big[U^g(k)\big]^\dagger_{\alpha,\beta} c_{\hat gk,\beta},~~~\forall~g\in X.
\eea
In the superconducting state, in order for the BdG Hamiltonian to be invariant under symmetry transformations, each crystalline symmetry operation $g\in X$ is dressed by a normal-state spin rotation $s_g\in S_0$ and a charge rotation $e^{\imth\phi_g\hat F}\in U(1)$:
\bea
g^\prime H_\text{BdG}(g^\prime)^{-1}=H_\text{BdG};~~~g^\prime=e^{\imth\phi_g\hat F}s_g\cdot g,~~~s_g\in S_0,~\forall~g\in X.
\eea
satisfying
\bea
g^\prime\cdot h^\prime=s(g,h)\cdot (gh)^\prime,~~~s(g,h)\in S_f.
\eea
The full symmetry group $G_f$ of the superconducting state (and of the BdG Hamiltonian) can be denoted by
\bea
G_f=\{(s,g)\equiv s\cdot g^\prime|g^\prime=e^{\imth\phi_g\hat F}s_g\cdot g,s\in S_f,g\in X\}
\eea
which is exactly captured by the group extension described in (\ref{central extension:space group}). The classification of different fermion symmetry groups $G_f$ is given by the classification of group extension in (\ref{central extension:space group}), which in general is a difficult mathematical problem. In the special case when $S_f$ is an Abelian group, the classification is given by 2nd cohomology group $\mathcal{H}^2_{[\rho]}(X,S_f)$. 

In terms of group representations, the electrons in the BdG Hamiltonian form a projective representation of crystalline group $X$:
\bea
\hat g^\prime c_{k,\alpha}(\hat g^\prime)^{-1}=\big[\tilde U^g(k)\big]^\dagger_{\alpha,\beta} c_{\hat gk,\beta},~~~\tilde U^g(k)=e^{\imth\phi_g}U^{s_g}U^g(k),~~~U^{s_g}\in SU(2),~s_g\in S_0.
\eea
satisfying
\bea
\tilde U^g(\hat h k)\tilde U^h(k)=\tilde U^{s(g,h)}\tilde U^{gh}(k),~~~s(g,h)\in S_f.
\eea
Equivalently $\{\tilde U^s\cdot\tilde U^g|s\in S_f,g\in X\}$ can be viewed as a linear representation of the full symmetry group $G_f$. 

Finally, we discuss the relation between the fermion PSG $G_f$ and the pairing symmetry. In general, the pairing wavefunctions $\Delta_{\alpha,\beta}$ in BdG Hamiltonian (\ref{ham:BdG}) form a linear representation $\mathcal{R}_{pair}$ of the bosonic symmetry group $G=S\times X$, where $S$ stands for the global (spin rotational) symmetry group and $X$ stands for the crystalline symmetry group. Meanwhile, in the representation $\{\tilde U^s\cdot\tilde U^g|s\in S_f,g\in X\}$ introduced above, we can identify a projective representation of group $G=S\times X$:
\bea
\mathcal{R}_\Phi(s,g)=e^{\imth(\phi_s+\phi_g)}U^{s_g},~~~\forall ~s\in S,~g\in X.
\eea
The linear representation $\mathcal{R}_{pair}$ which characterizes the pairing symmetry and the above projective representation $\mathcal{R}_\Phi$ are related by the following relation:
\bea\label{Rep pair:general}
\mathcal{R}_\Phi\otimes\mathcal{R}_\Phi\otimes \mathcal{R}_{pair}=\mathbbm{1}\oplus\cdots
\eea
where $\mathbbm{1}$ denotes the trivial one-dimensional (1d) representation of group $G=S\times X$. This is because the pairing term (\ref{ham:pairing}) must remain invariant under the PSG symmetry transformation $\{\tilde U^s\cdot\tilde U^g|s\in S_f,g\in X\}$. Notice that in the special case of $\mathcal{R}_\Phi$ being a 1d irrep, applicable to the situation discussed in Section \ref{sec:PSG-->pairing sym}, the general relation (\ref{Rep pair:general}) reduces to Eq. (\ref{Rep pair:1d}). Specifically, while the pairing order parameter $\Delta_{\alpha,\beta}$ always transforms as a 1d irrep of the global symmetry group $S$, it can transform as a multi-dimensional irrep. of the crystalline group $X$. 

Below we demonstrate the above general framework in two famous examples: the A and B phases of superfluid Helium 3. In the example of superfluid B phase, the pairing wavefunctions transform as a 3-dimensional odd-parity $j=1$ representation of $X=O(3)$ group, as will be shown below.  

\subsection{Examples: superfluid A and B phases in Helium 3}\label{sec:helium 3}

The most famous example of triplet superconductivity (or superfluidity) is perhaps Helium 3 \cite{leggett1975theoretical}, where the normal state preserves both the continuous $X_0=O(3)=SO(3)_L\times Z_2^I$ spatial symmetry and $S_0=SO(3)_S$ spin rotational symmetry. The full spatial symmetry $X=O(3)$ includes both spatial $SO(3)_L$ rotations, and inversion symmetry $Z_2^I=\{\hat I,1=(\hat I)^2\}$. There are two different superconducting phases in Helium 3: (1) B phase, also known as the Balian-Werthamer (BW) phase \cite{balian1963superconductivity}, (2) A phase, also known as Anderson-Brinkman-Morel (ABM) phase \cite{anderson1961generalized,andersonbrinkman1973}. Below we apply the general framework described above to these two triplet superconducting phases. The BdG Hamiltonian for superfluid Helium 3 can be generally written as
\bea\label{BdG ham:Helium 3}
\hat H_\text{BdG}=\sum_k\Psi_k^\dagger\bpm(\frac{k^2}{2m}-\mu)\hat 1&\vec d_k\cdot\vec\sigma\\(\vec d_k)^\ast\cdot\vec\sigma&(\mu-\frac{k^2}{2m})\hat 1\epm\Psi_k
\eea
in the Nambu basis of
\bea
\Psi_k\equiv(c_{k,\uparrow},c_{k,\downarrow},c^\dagger_{-k,\downarrow},-c^\dagger_{-k,\uparrow})^T.
\eea
The complex vector $\vec d_k\in\mathbb{C}^3$ determines the pairing symmetry of the system. 

In the normal state, the electron operators $c_{k,\uparrow/\downarrow}$ form a spinful representation of $SO(3)_S$ spin rotational symmetry
\bea\label{3He:normal state spin rotation}
U^{r_{\hat n}(\theta)}=e^{\imth\theta\hat n\cdot\vec\sigma/2}
\eea
where $r_{\hat n}(\theta)$ is the spin rotation by angle $\theta$ along axis $\hat n$. Under spatial rotation $R_{\hat n}(\theta)$ of angle $\theta$ along $\hat n$ axis, and under spatial inversion $\hat I$, the electrons in the normal state transform as
\bea
U^{R_{\hat n}(\theta)}=U^{I}=\hat 1.
\eea

\subsubsection{Superfluid B phase of Helium-3}\label{sec:He3:B phase}

The superfluid B phase preserves the full $X=O(3)=SO(3)_L\times Z_2$ group of spatial rotation and inversion symmetry, while the spin rotational symmetry $S_0=SO(3)_S$ is completely broken down to a trivial group $S=1$. According to central extension (\ref{central extension:onsite sym}), the fermion onsite symmetry group is simply the fermion parity group $S_f=Z_2^F$. Therefore the full fermion symmetry group $G_f$ given by central extension (\ref{central extension:space group}) is classified by
\bea
\mathcal{H}^2(X,S_f)=\mathcal{H}^2(SO(3)_L\times Z_2^I,Z_2^F)=\mathcal{H}^2(SO(3)_L,Z_2^F)\times\mathcal{H}^2(Z_2^I,Z_2^F)=\mbz_2\times \mbz_2
\eea
The superfluid B phase corresponds to the nontrivial extension of both $SO(3)_L$ spatial rotations and inversion $Z_2$, i.e. the nontrival element of both $\mbz_2$ subgroup of the $\mbz_2^2$ classification, featuring
\bea
\vec d_k\propto\vec k
\eea
Specifically, denoting an arbitrary element of $SO(3)_L$ rotational group as $(\hat n,\theta)$ where $\hat n$ is a unit vector of rotation axis and $\theta\in[0,2\pi)$ is the rotation angle, the $X=SO(3)_L$ spatial rotation symmetry in (\ref{BdG ham:Helium 3}) is implemented by 
\bea
\tilde U^{R_{\hat n}(\theta)}= e^{\imth\theta\hat n\cdot\vec\sigma/2}=\mathcal{R}_\Phi(R_{\hat n}(\theta))
\eea
in the superfluid B phase. Note that in the normal state $\{U^g_0\equiv\mathbbm{1}|g\in SO(3)_L\}$ form a $l=0$ representation of $SO(3)$ group, which is a linear representation. In the superfluid B phase, $\{\tilde U^g|g\in SO(3)_L\}$ form a $j=1/2$ representation of $SO(3)_L$ group, which is a projective representation. According to relation (\ref{Rep pair:general}) and the angular momentum addition rules, the pairing wavefunctions $\Delta_{\alpha,\beta}$ form either a $j=0$ or $j=1$ irrep of $SO(3)_L$ group. However, because the projective representation $\mathcal{R}_\Phi(R_{\hat n}(\theta))$ coincides with the normal-state spin rotation $U^{r_{\hat n}(\theta)}$ in (\ref{3He:normal state spin rotation}), the $j=0$ irrep will preserve the $SO(3)_S$ spin rotational symmetry, and hence does not applied to the superfluid B phase where $SO(3)_S$ is spontaneously broken. As a result, the pairing order parameters $\vec d_k\propto\vec k$ must form a 3-dimensional $j=1$ representation of the $SO(3)_L$ group of spatial rotations. 

Considering the inversion symmetry $\hat I$, the electrons transform as
\bea
\tilde U^{ I}=\imth\hat 1,~~~\hat I^\prime=e^{\imth\frac\pi2\hat F}\hat I,~~~\mathcal{R}_\Phi(I)=e^{\imth\pi/2}\mathbbm{1}
\eea
in superfluid B phase, with a nontrivial 2-cocycle
\bea
\hat I^\prime\times\hat I^\prime=s(\hat I,\hat I)=(-1)^{\hat F}
\eea
This corresponds to a pairing order parameter of odd inversion parity:
\bea
\vec d_{-k}=-\vec d_k
\eea
Therefore, the fermion symmetry group is $G_f=SU(2)\times Z_4^{I^\prime}/Z_2$ which is a nontrivial central extension of $X=O(3)$.

\subsubsection{Superfluid A phase of Helium-3}\label{sec:He3:A phase}

The superfluid A phase features
\bea
\vec d_k=\vec k\cdot(\hat m+\imth\hat n)(\hat m\times \hat n),~~~\hat m\perp\hat n.
\eea
where $\hat m,\hat n$ are two perpendicular unit vectors. Without loss the generality, the usual convention chooses $\hat m=\hat x$ and $\hat n=\hat y$, and we can write the $\vec d_k$ vector as
\bea
\vec d_k=(k_x+\imth k_y)\hat z
\eea
The spin rotational symmetry is broken from $S_0=SO(3)$ down to $S=O(2)=U(1)_{S^z}\rtimes Z_2^x$, generated by $U(1)_{r_z}=\{r_z(\theta)\equiv e^{\imth\theta S^z}|0\leq\theta<2\pi\}$ spin rotations along $\hat z=\hat m\times\hat n$ axis, and 2-fold spin rotation $r_x(\pi)\equiv e^{\imth\pi S^x}$ along $\hat x=\hat m$ axis. All possible fermion onsite symmetry groups $S_f$ are classified by 2nd cohomology group:
\bea
\mathcal{H}^2(S,Z_2^F)=\mathcal{H}^2(O(2),Z_2)=\mbz_2^3
\eea
In the superfluid A phase, spin rotational symmetries are implemented by
\bea
&\tilde U^{r_x(\pi)}=-\imth e^{\imth\frac\pi2\sigma_x}=\sigma_x,~~~\tilde U^{r_z(\theta)}=e^{\imth\theta\sigma_z/2}
\eea
associated with a nontrivial extension of $O(2)$ group with the following 2-cocycle:
\bea
\omega(r_x(\pi),r_x(\pi))=+1,~~\omega(r_z(\pi),r_z(\pi))=\omega(r_y(\pi),r_y(\pi))=-1.
\eea
This corresponds to a fermion global symmetry of $S_f\simeq O(2)$. 

The spatial $O(3)$ symmetry is broken down to a subgroup of $X=U(1)_{R_z}\times Z_2^I$, generated by $U(1)_{R_z}=\{R_z(\theta)|0\leq\theta<2\pi\}$ spatial rotations and spatial inversion $\hat I$. In the superfluid A phase, they are represented by
\bea
\tilde U^{R_z(\theta)}=e^{-\imth\theta/2}\hat 1,~~~\tilde U^{I}=\imth\hat 1.
\eea
In this case the fermion symmetry group $G_f$ is a nontrivial extension of $X=U(1)_{R_z}\times Z_2^I$ by $S_f=O(2)$
\bea
1\rightarrow O(2)\rightarrow G_f\rightarrow U(1)_{R_z}\times Z_2^I\rightarrow 1
\eea
satisfying
\bea
\big[R_z^\prime(\pi)\big]^2=(\hat I^\prime)^2=(-1)^{\hat F}
\eea
The associated fermion PSG is $G_f=O(2)\times U(1)\times Z_4^{I^\prime}/Z_2$. 


\section{Conclusions and outlook}\label{sec:conclusion}

Traditionally, the broken and unbroken symmetries of a superconductor (SC) is described by the Ginzburg-Landau theory, which characterizes the symmetry properties of all bosonic excitations therein, such as Cooper pairs. In this paper we investigate the same problem of broken and unbroken symmetries in a SC state from a viewpoint of fermionic excitations. We showed that the projective symmetry group (PSG) of fermions in a superconductor is the proper language to capture symmetry-related properties of fermionic excitations in a SC, and systematically studied the relationship between the pairing symmetry and the fermion PSGs in a superconductor. We provided a general framework in Section \ref{sec:general} to characterize the fermion symmetry group after the Cooper pair formation with the concept of PSG, which is a group extension of the crystalline space group $X$ by the fermion global symmetry group $S_f$ in the superconducting phase. Examples of fermion global symmetry groups include the fermion parity group $Z_2^F$ in a generic SC without spontaneous breaking of spin rotational symmetries, and $O(2)$ as in the case of superfluid A phase of Helium-3. In the case of the fermion global symmetry group $S_f$ being an Abelian group, we encapsulated the group extension problem in the language of second group cohomology, which is both conceptually clear and practically easy to tackle. 

When the SC and normal state share the same fermion global symmetries, i.e. in the absence of spontaneously broken global (spin rotational) symmetries, the fermion PSG of the SC state is particularly simple: it is a central extension of the crystalline symmetry group $X$ by the fermion parity group $Z_2^F$. In this case, we can classify all fermion PSGs using elements of the 2nd cohomology group $\mathcal{H}^2(X,Z_2^F)$. Using the connection between pairing symmetry and fermion PSG discussed in section \ref{sec:psg}, we can systematically obtain all the possible pairing symmetries compatible with the PSGs as delineated in Sec. \ref{sec:PSG-->pairing sym}. A distinction was made between the case of SCs with and without spin-orbital couplings (SOCs), where in the presence of a strong SOC, crystalline symmetries of fermions in the normal state are described by a non-trivial 2-cocycle $\omega_0\in \mathcal{H}^2(X,Z_2^F)$, and the correspondence between PSG and pairing symmetry should be shifted accordingly.  Within this general framework, we calculated all the possible PSGs for all 3-dimensional point group symmetries both with and without SOCs, and establish the correspondence between PSGs and pairing symmetries of the SCs. As a demonstration of the framework, we studied in detail the PSGs and pairing symmetries of several physically relevant systems in section \ref{sec:examples}, and hope our work would shed new lights on understandings of superconductivity in these systems. Considering the crystalline symmetry group $X$, although we have restricted our attention to point groups in this work, the case of magnetic point groups and space groups can be naturally incorporated in our general framework.

PSGs have important implications on physical properties of a superconductor. As the PSG $G_f$ is the symmetry group of fermions in a SC, it dictates the symmetry and topological properties of all the fermionic excitations of the system and its validity extends beyond the mean-field BdG equations. Therefore, PSG can be used to classify topological superconductors in both non-interacting (i.e., admitting a mean-field description) and interacting cases. As an illustration, we discussed systems with two different kinds of symmetry groups where $G_f$ determines classifications of non-interacting topological superconductors. Moreover, as PSG establishes a link between pairing symmetry and topological properties of a system, we can utilize topological properties of the electronic excitations as a diagnosis for the pairing symmetry of a superconductor. We leave these interesting ideas for future works.

\acknowledgements{This work is supported by Center for Emergent Materials at The Ohio State University, a National Science Foundation (NSF) MRSEC through NSF Award No. DMR-2011876. YML acknowledges support by grant NSF PHY-1748958 to the Kavli Institute for Theoretical Physics (KITP), and NSF PHY-2210452 to the Aspen Center for Physics.}

\appendix

\section{A short introduction to projective representation and 2-cocycle}\label{app:psg}

In this appendix we want to elucidate the connection between the projective representation of the crystalline symmetry group as described by the mathematical object called 2-cocycle and the fermion projective symmetry group $G_f$.

The concept of PSG was first introduced in the study of quantum spin liquids\cite{Wen2002}. In the context of quantum spin liquids, electrons can be thought of as being composed of chargons and spinons which are glued together by an $SU(2)$ gauge field. Due to the emergent gauge structures, symmetries that are represented linearly on the physical degrees of freedom are now represented only projectively on the spinons. More specifically, spin operators at site $i$ can be written as fermionic spinons: $S_i=\frac{1}{2}f_{i,\alpha}^{\dagger}\vec{\sigma}_{\alpha,\beta}f_{i,\beta}$. A spin Hamiltonian can be described by a mean-field theory of spinons plus gauge fluctuations. Consider the following mean-field Hamiltonian: 
\bea
H=\sum\limits_{ij}[\psi_i^{\dagger}u_{ij}\psi_j+h.c.]+\sum\limits_ia_0^l\psi_i^{\dagger}\tau^l\psi_i,
\eea
where $u_{ij}$'s are $2\times 2$ matrices encoding pairing and hoppings of fermionic spinons, $\psi_i=(f_{\uparrow},f_{\downarrow}^{\dagger})^T$ are Nambu spinors. 

The Hamiltonian has a local $SU(2)$ gauge redundancy: a site-dependent $SU(2)$ transformation $\psi_i\rightarrow W_i\psi_i, u_{ij}\rightarrow W_iu_{ij}W_j^{\dagger}$ with $W_i\in SU(2) $ which leaves both physical observables and the Hamiltonian invariant. Due to this gauge redundancy, the symmetry of the spin liquids are described by the projective symmetry group, which is defined as the collection of all combinations of symmetry elements and gauge transformations that leave the mean-field ansatz $\{u_{ij}\}$ invariant: 
\bea
&G_UU(\{u_{ij}\})=\{u_{ij}\},\\
&U(\{u_{ij}\}\equiv \{\tilde{u}_{ij}=u_{U^{-1}(i),U^{-1}(j)}\},\\
&G_U(u_{ij})\equiv\{\tilde{u}_{ij}=G_{U}(i)u_{ij}G_{U}^{\dagger}(j)\},\\
&G_{U}(i)\in \text{SU}(2),
\eea
where $U$ is an element of the symmetry group $SG$ of the microscopic system and $G_U$ is the $SU(2)$ gauge transformation accompanying $U$ that leaves the mean-field ansatz invariant.

To encode the emergent gauge fields at low energy for spin liquid states, we introduce the important concept of invariant gauge group (IGG) which are pure gauge group elements that leave the mean-field ansatz invariant: $ W_iu_{ij}W_{j}^{\dagger}=u_{ij}$. It is clear that IGG corresponds to elements $G_UU$ in PSG where $U$ is the identity. With the concept of IGG it is now easy to describe the structure of PSG. In fact, IGG is a normal subgroup of PSG, and with the group homomorphism $\rho(G_UU)=U$ between PSG and SG, we have the following exact sequence:
\bea\label{short exact sequence}
1\rightarrow \text{IGG}\xrightarrow{\iota} \text{PSG}\xrightarrow{\rho}
\text{SG}\rightarrow 1,
\eea
where $\iota$ is the embedding mapping, and the exactness is ensured by the fact that $\rho(w)\equiv \textbf{1}\in$ SG for $w\in$ IGG. The structure of the PSG is now quite clear: it is the group extension of the SG by the IGG, or alternatively, SG$=$PSG/IGG. 

Equipped with the knowledge of PSG, it is also easy to see that the problem of unbroken symmetries of the superconductor naturally fits into the general framework of PSG if we notice that the BdG Hamiltonian takes the same form as the spin liquid mean-field Hamiltonian. More precisely, as discussed in the main text, fermions in the superconductor has the symmetry group $G_f$, which is an extension of the space group $X$ by the fermion global symmetry group $S_f$ described by the short exact sequence:
\bea
1\rightarrow S_f\rightarrow G_f\rightarrow X\rightarrow 1.
\eea
The resemblance to Eq.~\ref{short exact sequence} is immediately seen if we identify the unbroken global symmetry group $S_f$ as IGG and the fermion symmetry group $G_f$ as PSG. However, there's an important difference we need to keep in mind: in our study of superconductor, the global symmetry group $S_f$ should not be regarded as the gauge group corresponding to a fluctuating gauge field, as was in the context of spin liquids.

In general $S_f$ can be non-Abelian, and we refer to Ref. \cite{Wen2002} for a general computation scheme to solve the extension problem by obtaining all the inequivalent projective symmetry groups $G_f$.
Below let's discuss the special case of $S_f$ being Abelian, which covers most of the practical situations and is mathematically much simpler to deal with. And we will comment briefly on the case of $S_f$ being non-Abelian in the end.

In the case of $S_f$ being Abelian, the group extension of $X$ by $S_f$ can be described as an element in the second cohomology group $\mathcal{H}^2_{[\rho]}(X,S_f)$ with group actions $[\rho]:X\rightarrow \text{Aut}(S_f)$ (note that when the group action is trivial, the group extension is simply a central extension). To see this more clearly, let's label group elements in $G_f$ as $(s,g)$ with $s\in S_f, g\in X$. Now, since $S_f$ is in the center of $G_f$, we can represent the group multiplication rule in the following way:
\bea
(s_g,g)\times (s_h,h)=(s(g,h)s_gs_h,gh),
\eea
where $s(g,h)$ is a function $X\times X\rightarrow S$. The above procedure has an ambiguity since we can alternatively define $g'=\gamma_gg\in G_f$ ($\gamma_g\in S_f$) as our canonical choice of $g$. This then modifies $s(g,h)$ as: 
\bea\label{eq:groupcoboundary}
s(g,h)\rightarrow s(g,h)\cdot\gamma_g\cdot\gamma^g_h\cdot\gamma_{gh}^{-1},
\eea
where the superscript $g$ indicates group actions $g$ on elements in $S_f$ as described by $[\rho]$ .

The $s(g,h)$'s satisfy the associativity condition if we apply three group elements in $G_f$ in two equivalent ways, which yields
\bea\label{eq:groupcocycle}
s(g_1,g_2)s(g_1g_2,g_3)=s(g_1,g_2g_3)s^{g_1}(g_2,g_3).
\eea

The coboundary condition \ref{eq:groupcoboundary} and the cocycle condition \ref{eq:groupcocycle} then define an element in $\mathcal{H}^2(X,S_f)$. Therefore we have found out that in the case of central extension, $G_f$ is uniquely determined by the 2-cocycle $s(g,h)$, which is further classified by the second cohomology group $\mathcal{H}^2(X,S_f)$.

Before proceeding, let me emphasize an important point: fermions fulfill a 1d representation of $S_f$, which we denote as $\rho_S:S_f\rightarrow U(1)$. Note that $\rho_S$ is determined by the microscopic electrons and can be viewed as a group homomorphism from $S_f$ to $\text{Image}(\rho_S)$.

Because elements in $G_f$ act on fermions in a linear way, let's consider a linear representation $\hat{U}$ of the group $G_f$. Since $S_f$ lies at the center of the group, $\hat{U}((s,1))$ should be of the form $\rho_S(s)\times \mathbbm{1}$ according to Schur's lemma and the fact that the symmetry action of $s\in S$ on fermions is given by $\rho_S$.

If we identify $U(g)$ as $\hat{U}((1,g))$, $U(g)$ would fulfill a projective representation of $X$: 
\bea\label{eq:correspondence}
U(g)U(h)=\hat{U}((1,g))\hat{U}((1,h))=\hat{U}((s(g,h),gh))=\hat{U}((s(g,h),1))\hat{U}((1,gh))=\omega(g,h)U(gh),
\eea
where $\omega(g,h)\equiv \rho_S(s(g,h))$ is a function $X\times X\rightarrow \text{Image}(\rho_S)$. The $\omega$ satisfies the following associativity condition if we act three consecutive symmetry operations in two equivalent ways: $g_1g_2g_3=(g_1g_2)g_3=g_1(g_2g_3)$, which translates to
\bea\label{eq:omegacocycle}
\omega(g_1,g_2)\omega(g_1g_2,g_3)=\omega(g_1,g_2g_3)\omega^{g_1}(g_2,g_3),
\eea
where the superscript $g$ on $\omega$ indicates group actions on the $U(1)$ phase induced by the group action $[\rho]$ on elements in $S_f$.

We can also multiply symmetry actions $U(g)$ by some $U(1)$ phase $\gamma_g\in\text{Image}(\rho_S)$, which then modifies $\omega$ in the following way:
\bea\label{eq:omegacoboundary}
\omega(g,h)\rightarrow \omega(g,h)\frac{\gamma_g\gamma^g_h}{\gamma_{gh}}.
\eea

The associativity condition (\ref{eq:omegacocycle}) and the ambiguity (\ref{eq:omegacoboundary}) thus define a 2-cocycle in the second cohomology group $\mathcal{H}^2(X,\text{Image}(\rho_S))$. And the equation Eq.\eqref{eq:correspondence} establishes an explicit homomorphism between the projective representation of $X$ (an element in $\mathcal{H}^2(X,\text{Image}(\rho_S))$) and the fermion projective symmetry group $G_f$ (an element in $\mathcal{H}^2(X,S)$).

In summary, a linear representation of the fermion projective symmetry group $G_f$ can alternatively be viewed as a projective representation of the group $X$ with cocycle $\omega(g,h)\in \mathcal{H}^2(X,\text{Image}(\rho_S))$, as elucidated by Eq.\eqref{eq:correspondence}.

Several remarks are in order:
\begin{enumerate}
    \item When $S_f$ is Abelian and $\rho_S$ is injective, the two cohomology groups $\mathcal{H}^2(X,\text{Image}(\rho_S))$ and $\mathcal{H}^2(X,S_f)$ are isomorphic to each other, therefore we have sometimes used these terms interchangeably in the main text.
    \item When $S_f$ is non-Abelian, $G_f$ can no longer be described by an element in the second cohomology group. If we restrict our attention to the case where the representation $\rho_S$ of $S_f$ on fermions are one dimensional, then the correspondence Eq.\eqref{eq:correspondence} still holds, enabling us to carry out calculations within this general framework.
    \item When $S_f$ is non-Abelian, there are cases where the representation of $S_f$ on fermions are at least 2-dimensional, such as spin-$1/2$ fermions in the superfluid A phase with $S_f=O(2)$. Such cases are beyond the scope of cohomological description, and we need to solve the projective symmetry groups up to gauge equivalence on a case-by-case basis following the general procedures as described in Ref.\cite{Wen2002}. 
\end{enumerate}

\section{How PSG constrains the pairing symmetry for all crystalline point groups}\label{app:table}

Since $G_f$ is the extension of $G$ by $Z_2^F$, we can view 1d projective representations $\mathcal{R}_{\Phi}(g)$ of $G$ as regular representations $\bar{\mathcal{R}}_{\Phi}(\hat{g}')$ for $\hat{g}'\in G_f$ with $\bar{\mathcal{R}}_{\Phi}(d)=-1$ ( $d\equiv (-1)^{\hat{F}}$) when restricted to the subgroup $X=G_f/Z_2^F$. This is confirmed by the following relation:
\bea
\bar{\mathcal{R}}_{\Phi}((\eta_g,\hat g^\prime) )\bar{\mathcal{R}}_{\Phi}((\eta_h,\hat h^\prime))=\bar{\mathcal{R}}_{\Phi}((\eta_g\eta_h\tilde\omega(g,h),\hat g^\prime\hat h^\prime))=\omega(g,h)\bar{\mathcal{R}}_{\Phi}((\eta_g\eta_h,\hat g^\prime\hat h^\prime)),~~~\forall~\eta_{g},\eta_h=\pm1,
\eea
where we have used the fact that $Z_2^F$ is the center of $G_f$.

Our strategy then is to first obtain the group extension $G_f\in \mathcal{H}^2(X,Z_2^F)$ and then compute the 1d irreducible representations $\bar{\mathcal{R}}_{\Phi}(g)$ of $G_f$ with $Z_2^F=-1$, from which we can readily obtain $\mathcal{R}_{pair}$. We used GAP computer algebra program \cite{GAP4} in all these calculations, which is ideally suited for the task. The results are displayed in Table.\ref{table:point_group}.

\newpage

\begin{longtable}{|c|c|c|c|c|c|c|c|}
       \caption{Correspondence between the fermion PSG and the representation of the pairing order parameter for all the crystalline point group. We list gauge-invariant cocycles to label different projective symmetry groups $G_f$ for superconductors both without and with spin-orbital couplings. We follow the convention in Ref.\cite{bradley2010} to label irreducible representations $\mathcal{R}_{pair}(g)$ of the pairing order parameter. Some $G_f$ does not admit a 1d projective representation and hence the corresponding $\mathcal{R}_{pair}$ is marked as N/A.}\label{table:point_group}
       \endfirsthead
       \caption*{\textbf{Table \ref{table:point_group}} Continued.}
       \endhead
    \hline
       $X$  &$\mathcal{H}^2(X,Z_2^F)$& Gauge-invariant 2-cocycles $\tilde\omega$  &No SOC (spinless)&w/ SOC (spinful)&$\mathcal{R}_{pair}(g)$ \\
       \hline
$C_1$ & $\mathbb{Z}_1$& $-$&$-$ &$-$&$A$ \\
       \hline
     \multirow{2}{*}{$C_i$} & \multirow{2}{*}{$\mbz_2$} &\multirow{2}{*}{$\omega(i,i)$}& 
$1$ & 
$1$& $A_g$\\ 
\cline{4-6}
&&&$-1$&$-1$&$A_u$\\ 
       \hline  
       \multirow{2}{*}{$C_2$} & \multirow{2}{*}{$\mbz_2$}& \multirow{2}{*}{$\omega(C_2,C_2)$} & 
$1$ &$-1$& $A$\\ 
\cline{4-6}
&&&$-1$&$1$&$B$\\ 
       \hline  
       
       \multirow{2}{*}{$C_s$} & \multirow{2}{*}{$\mbz_2$}&\multirow{2}{*}{$\omega(\sigma_h,\sigma_h)$} & 
$1$ & 
$-1$ & $A'$\\ 
\cline{4-6}
&&&$-1$& 
$1$&$A''$\\ 
       \hline  

   \multirow{5}{*}{$C_{2h}$} & \multirow{5}{*}{$\mbz_2^3$} & \multirow{5}{*}{$(\omega(C_2,C_2),\omega(i,i),\omega(\sigma_h,\sigma_h))$}&
$(1,1,1)$&
$(-1,1,-1)$ & $A_g$\\
\cline{4-6}
&&&$(1,-1,-1)$&
$(-1,-1,1)$ &$A_u$\\ 
\cline{4-6}
&&&$(-1,1,-1)$&
$(1,1,1)$ &$B_g$\\ 
\cline{4-6}
&&&$(-1,-1,1)$&
$(1,-1,-1)$ &$B_u$\\ 
\cline{4-6}
&&&other cases&other cases&N/A\\ 
\cline{4-6}
       \hline  

\multirow{5}{*}{$D_{2}$} & \multirow{5}{*}{$\mbz_2^3$}& \multirow{5}{*}{$(\omega(C_{2x},C_{2x}),\omega(C_{2y},C_{2y}),\omega(C_{2z},C_{2z}))$} & 
$(1,1,1)$ & 
$(-1,-1,-1)$& $A$\\
\cline{4-6}
&&&$(-1,-1,1)$ & 
$(1,1,-1)$&$B_1$\\ 
\cline{4-6}
&&&$(-1,1,-1)$& 
$(1,-1,1)$ &$B_2$\\ 
\cline{4-6}
&&&$(1,-1,-1)$& 
$(-1,1,1)$ &$B_3$\\ 
\cline{4-6}
&&&other cases&other cases&N/A\\ 
\cline{4-6}
       \hline  

       \multirow{5}{*}{$C_{2v}$} & \multirow{5}{*}{$\mbz_2^3$} & \multirow{5}{*}{$(\omega(C_2,C_{2}),\omega(\sigma_v,\sigma_v),\omega(\sigma_v',\sigma_v'))$} & 
$(1,1,1)$& 
$(-1,-1,-1)$ & $A_1$\\
\cline{4-6}
&&&$(1,-1,-1)$& 
$(-1,1,1)$ &$A_2$\\ 
\cline{4-6}
&&&$(-1,1,-1)$& 
$(1,-1,1)$ &$B_1$\\ 
\cline{4-6}
&&&$(-1,-1,1)$& 
$(1,1,-1)$ &$B_2$\\ 
\cline{4-6}
&&&other cases&other cases&N/A\\ 
\cline{4-6}
       \hline  

  \multirow{9}{*}{$D_{2h}$} & \multirow{9}{*}{$\mbz_2^6$}
  & \multirow{9}{*}{$(\omega(C_{2x},C_{2x}),\omega(C_{2y},C_{2y}),\omega(i,i),\frac{\omega(C_{2x},i)}{\omega(i,C_{2x})},\frac{\omega(C_{2y},i)}{\omega(i,C_{2y})},\frac{\omega(C_{2x},C_{2y})}{\omega(C_{2y},C_{2x})})$}
  & 
$(1,1,1,1,1,1)$& 
$(-1,-1,1,1,1,-1)$ & $A_g$\\
\cline{4-6}
&&&$(1,-1,1,1,1,1)$& 
$(-1,1,1,1,1,-1)$ &$B_{3g}$\\ 
\cline{4-6}
&&&$(-1,1,1,1,1,1)$& 
$(1,-1,1,1,1,-1)$ &$B_{2g}$\\ 
\cline{4-6}
&&&$(-1,-1,1,1,1,1)$& 
$(1,1,1,1,1,-1)$ &$B_{1g}$\\ 
\cline{4-6}
&&&$(1,1,-1,1,1,1)$& 
$(-1,-1,-1,1,1,-1)$ &$A_u$\\ 
\cline{4-6}
&&&$(1,-1,-1,1,1,1)$& 
$(-1,1,-1,1,1,-1)$ &$B_{3u}$\\ 
\cline{4-6}
&&&$(-1,1,-1,1,1,1)$& 
$(1,-1,-1,1,1,-1)$ &$B_{2u}$\\ 
\cline{4-6}
&&&$(-1,-1,-1,1,1,1)$ & 
$(1,1,-1,1,1,-1)$&$B_{1u}$\\ 
\cline{4-6}
&&&other cases&other cases&N/A\\ 
\cline{4-6}
       \hline

\multirow{2}{*}{$C_4$} & \multirow{2}{*}{$\mbz_2$}&\multirow{2}{*}{$\omega(C_2,C_2)$} & 
$1$  & 
$-1$ & $A,B$\\ 
\cline{4-6}
&&&$-1$ & 
$1$ &$E$\\ 
       \hline  

       \multirow{2}{*}{$S_4$} & \multirow{2}{*}{$\mbz_2$}&\multirow{2}{*}{$\omega(C_2,C_2)$} & 
$1$  & 
$-1$ & $A,B$\\ 
\cline{4-6}
&&&$-1$ & 
$1$ &$E$\\ 
       \hline

        \multirow{5}{*}{$C_{4h}$} & \multirow{5}{*}{$\mbz_2^3$} & \multirow{5}{*}{$(\omega(C_2,C_{2}),\omega(i,i),\frac{\omega(C_4,i)}{\omega(i,C_4)})$}& 
$(1,1,1)$ & 
$(-1,1,1)$& $A_g,B_g$\\
\cline{4-6}
&&&$(-1,-1,1)$  & 
$(1,-1,1)$&$E_u$\\ 
\cline{4-6}
&&&$(1,-1,1)$ & 
$(-1,-1,1)$ &$A_u,B_u$\\ 
\cline{4-6}
&&&$(-1,1,1)$ & 
$(1,1,1)$ &$E_g$\\ 
\cline{4-6}
&&&other cases&other cases&N/A\\ 
\cline{4-6}
       \hline

       \multirow{5}{*}{$D_4$} & \multirow{5}{*}{$\mbz_2^3$} & \multirow{5}{*}{$(\omega(C_2,C_{2}),\omega(C_2',C_2'),\omega(C_2'',C_2''))$}& 
$(1,1,1)$ & 
$(-1,-1,-1)$& $A_1$\\
\cline{4-6}
&&&$(1,-1,-1)$& 
$(-1,1,1)$ &$A_2$\\ 
\cline{4-6}
&&&$(1,1,-1)$& 
$(-1,-1,1)$ &$B_1$\\ 
\cline{4-6}
&&&$(1,-1,1)$& 
$(-1,1,-1)$ &$B_2$\\ 
\cline{4-6}
&&&other cases&other cases&N/A\\ 
\cline{4-6}
       \hline

        \multirow{5}{*}{$C_{4v}$} & \multirow{5}{*}{$\mbz_2^3$} & \multirow{5}{*}{$(\omega(C_2,C_{2}),\omega(\sigma_v,\sigma_v),\omega(\sigma_d,\sigma_d))$}&
$(1, 1, 1)$ & 
$(-1,-1,-1)$& $A_1$\\
\cline{4-6}
&&&$(1, 1, -1)$& 
$(-1,-1,1)$ &$B_1$\\ 
\cline{4-6}
&&&$(1, -1, 1)$& 
$(-1,1,-1)$ &$B_2$\\ 
\cline{4-6}
&&&$(1, -1, -1)$& 
$(-1,1,1)$ &$A_2$\\ 
\cline{4-6}
&&&other cases&other cases&N/A\\ 
\cline{4-6}
\hline

 \multirow{5}{*}{$D_{2d}$}& \multirow{5}{*}{$\mbz_2^3$}& \multirow{5}{*}{($\omega(C_2,C_2),\omega(C_2',C_2'),\omega(\sigma_d,\sigma_d)$)}  & 
 $(1,1,1)$& 
 $(-1,-1,-1)$
 & $A_1$\\
\cline{4-6}
&&& $(1,1,-1)$& 
 $(-1,-1,1)$ &$B_1$\\ 
\cline{4-6}
&&& $(1,-1,1)$& 
 $(-1,1,-1)$ &$B_2$\\ 
\cline{4-6}
&&& $(1,-1,-1)$& 
 $(-1,1,1)$ &$A_2$\\ 
\cline{4-6}
&&&other cases&other cases&N/A\\ 
\cline{4-6}
       \hline  

        \multirow{9}{*}{$D_{4h}$} & \multirow{9}{*}{$\mbz_2^6$} & \multirow{9}{*}{$(\omega(C_2',C_2'),\omega(C_2'',C_2''),\omega(i,i),\omega(C_2,C_2),\frac{\omega(C_2',i)}{\omega(i,C_2')},\frac{\omega(C_2'',i)}{\omega(i,C_2'')})$)}
        
        & 
$(1,1,1,1,1,1)$ & 
$(-1,-1,1,-1,1,1)$ & $A_{1g}$\\
\cline{4-6}
&&&$(1,1,-1,1,1,1)$& 
$(-1,-1,-1,-1,1,1)$ &$A_{1u}$\\ 
\cline{4-6}
&&&$(1,-1,1,1,1,1)$& 
$(-1,1,1,-1,1,1)$ &$B_{1g}$\\ 
\cline{4-6}
&&&$(1,-1,-1,1,1,1)$& 
$(-1,1,-1,-1,1,1)$ &$B_{1u}$\\ 
\cline{4-6}
&&&$(-1,-1,1,-1,1,1)$& 
$(1,1,1,1,1,1)$&$A_{2g}$\\ 
\cline{4-6}
&&&$(-1,-1,-1,1,1,1)$ & 
$(1,1,-1,-1,1,1)$&$A_{2u}$\\ 
\cline{4-6}
&&&$(-1,1,1,1,1,1)$& 
$(1,-1,1,-1,1,1)$ &$B_{2g}$\\ 
\cline{4-6}
&&&$(-1,1,-1,1,1,1)$& 
$(1,-1,-1,-1,1,1)$ &$B_{2u}$\\ 
\cline{4-6}
&&&other cases&other cases&N/A\\ 
\cline{4-6}
       \hline 
\pagebreak
\hline
$X$  &$\mathcal{H}^2(X,Z_2^F)$& Gauge-invariant cocycles   &No SOC&With SOC&$\mathcal{R}_{pair}(g)$ \\
       \hline

       $C_3$ &$\mathbb{Z}_1$&  $-$&$-$& $-$&$A_1,E$  \\
       \hline  

 \multirow{2}{*}{$C_{3i}$}& \multirow{2}{*}{$\mbz_2$}& \multirow{2}{*}{$\omega(i,i)$}  & 
 $+1$& 
 $+1$
 & $A_g,E_g$\\
\cline{4-6}
&&& $-1$& 
 $-1$ &$A_u,E_u$\\ 
\cline{4-6}
       \hline  

        \multirow{2}{*}{$D_{3}$}& \multirow{2}{*}{$\mbz_2$}& \multirow{2}{*}{$\omega(C_2,C_2)$}  & 
 $+1$& 
 $-1$
 & $A_1$\\
\cline{4-6}
&&& $-1$& 
 $+1$ &$A_2$\\ 
\cline{4-6}
       \hline

        \multirow{2}{*}{$C_{3v}$}& \multirow{2}{*}{$\mbz_2$}& \multirow{2}{*}{$\omega(\sigma_v,\sigma_v)$}  & 
 $+1$& 
 $-1$
 & $A_1$\\
\cline{4-6}
&&& $-1$& 
 $+1$ &$A_2$\\ 
\cline{4-6}
       \hline

 \multirow{5}{*}{$D_{3d}$}& \multirow{5}{*}{$\mbz_2^3$}& \multirow{5}{*}{($\omega(C_2',C_2'),\omega(i,i),\frac{\omega(C_2',i)}{\omega(i,C_2')}$)}  & 
 $(1,1,1)$& 
 $(-1,1,1)$
 & $A_{1g}$\\
\cline{4-6}
&&& $(1,-1,1)$& 
 $(-1,-1,1)$ &$A_{1u}$\\ 
\cline{4-6}
&&& $(-1,1,1)$& 
 $(1,1,1)$ &$A_{2g}$\\ 
\cline{4-6}
&&& $(-1,-1,1)$& 
 $(1,-1,1)$ &$A_{2u}$\\ 
\cline{4-6}
&&&other cases&other cases&N/A\\ 
\cline{4-6}
       \hline

      \multirow{2}{*}{$C_{6}$}& \multirow{2}{*}{$\mbz_2$}& \multirow{2}{*}{$\omega(C_2,C_2)$}  & 
 $+1$& 
 $-1$
 & $A,E_1$\\
\cline{4-6}
&&& $-1$& 
 $+1$ &$B,E_2$\\ 
\cline{4-6}
       \hline

      \multirow{2}{*}{$C_{3h}$}& \multirow{2}{*}{$\mbz_2$}& \multirow{2}{*}{$\omega(\sigma_h,\sigma_h)$}  & 
 $+1$ & 
 $-1$
 & $A',E'$\\
\cline{4-6}
&&& $-1$& $1$ &$A'',E''$\\ 
\cline{4-6}
       \hline 

      \multirow{5}{*}{$C_{6h}$}& \multirow{5}{*}{$\mbz_2^3$}& \multirow{5}{*}{($\omega(C_2,C_2),\omega(i,i),\frac{\omega(C_2,i)}{\omega(i,C_2)}$)}  & 
 $(1,1,1)$& 
 $(-1,1,1)$
 & $A_{g},E_{1g}$\\
\cline{4-6}
&&& $(1,-1,1)$& 
 $(-1,-1,1)$ &$A_{u},E_{1u}$\\ 
\cline{4-6}
&&& $(-1,1,1)$& 
 $(1,1,1)$ &$B_{g},E_{2g}$\\ 
\cline{4-6}
&&& $(-1,-1,1)$ & 
 $(1,-1,1)$&$B_{u},E_{2u}$\\ 
\cline{4-6}
&&&other cases&other cases&N/A\\ 
\cline{4-6}
       \hline 

   \multirow{5}{*}{$D_{6}$}& \multirow{5}{*}{$\mbz_2^3$}& \multirow{5}{*}{($\omega(C_2,C_2),\omega(C_2',C_2'),\frac{\omega(C_2,C_2')}{\omega(C_2',C_2)}$)}  & 
 $(1,1,1)$& 
 $(-1,-1,-1)$
 & $A_1$\\
\cline{4-6}
&&& $(1,-1,1)$& $(-1,1,-1)$ &$A_2$\\ 
\cline{4-6}
&&& $(-1,1,1)$& $(1,-1,-1)$ &$B_2$\\ 
\cline{4-6}
&&& $(-1,-1,1)$& $(1,1,-1)$ &$B_1$\\ 
\cline{4-6}
&&&other cases&other cases&N/A\\ 
\cline{4-6}
       \hline 
 
   \multirow{5}{*}{$C_{6v}$}& \multirow{5}{*}{$\mbz_2^3$}& \multirow{5}{*}{($\omega(C_2,C_2),\omega(\sigma_v,\sigma_v),\frac{\omega(C_2,\sigma_v)}{\omega(\sigma_v,C_2)}$)}  & 
 $(1,1,1)$& 
 $(-1,-1,-1)$
 & $A_1$\\
\cline{4-6}
&&& $(1,-1,1)$& 
 $(-1,1,-1)$ &$A_2$\\ 
\cline{4-6}
&&& $(-1,1,1)$& 
 $(1,-1,-1)$ &$B_2$\\ 
\cline{4-6}
&&& $(-1,-1,1)$& 
 $(1,1,-1)$ &$B_1$\\ 
\cline{4-6}
&&&other cases&other cases&N/A\\ 
\cline{4-6}
       \hline 

          \multirow{5}{*}{$D_{3h}$}& \multirow{5}{*}{$\mbz_2^3$}& \multirow{5}{*}{($\omega(\sigma_v,\sigma_v),\omega(\sigma_h,\sigma_h),\frac{\omega(\sigma_h,\sigma_v)}{\omega(\sigma_v,\sigma_h)}$)}  & 
 $(1,1,1)$& 
 $(-1,-1,-1)$
 & $A_1'$\\
\cline{4-6}
&&& $(1,-1,1)$& 
 $(-1,1,-1)$ &$A_2''$\\ 
\cline{4-6}
&&& $(-1,1,1)$& 
 $(1,-1,-1)$ &$A_2'$\\ 
\cline{4-6}
&&& $(-1,-1,1)$& 
 $(1,1,-1)$ &$A_1''$\\ 
\cline{4-6}
&&&other cases&other cases&N/A\\ 
\cline{4-6}
    \hline
\multirow{9}{*}{$D_{6h}$} & \multirow{9}{*}{$\mbz_2^6$} & \multirow{9}{*}{($\omega(C_2,C_2),\omega(C_2',C_2'),\omega(i,i),\frac{\omega(C_2,C_2')}{\omega(C_2',C_2)},\frac{\omega(C_2,i)}{\omega(i,C_2)},\frac{\omega(C_2',i)}{\omega(i,C_2')}$)}
        
        & 
$(1,1,1,1,1,1)$& 
$(-1,-1,1,-1,1,1)$ & $A_{1g}$\\
\cline{4-6}
&&&$(1,1,-1,1,1,1)$& 
$(-1,-1,-1,-1,1,1)$  &$A_{1u}$\\ 
\cline{4-6}
&&&$(1,-1,1,1,1,1)$& 
$(-1,1,1,-1,1,1)$  &$A_{2g}$\\ 
\cline{4-6}
&&&$(1,-1,-1,1,1,1)$& 
$(-1,1,-1,-1,1,1)$  &$A_{2u}$\\ 
\cline{4-6}
&&&$(-1,1,1,1,1,1)$& 
$(1,-1,1,-1,1,1)$ &$B_{2g}$\\ 
\cline{4-6}
&&&$(-1,1,-1,1,1,1)$& 
$(1,-1,-1,-1,1,1)$  &$B_{2u}$\\ 
\cline{4-6}
&&&$(-1,-1,1,1,1,1)$& 
$(1,1,1,-1,1,1)$  &$B_{1g}$\\ 
\cline{4-6}
&&&$(-1,-1,-1,1,1,1)$ & 
$(1,1,-1,-1,1,1)$ &$B_{1u}$\\ 
\cline{4-6}
&&&other cases&other cases&N/A\\ 
\cline{4-6}
       \hline 

      \multirow{2}{*}{$T$}& \multirow{2}{*}{$\mbz_2$}& \multirow{2}{*}{$\omega(C_2,C_2)$}  & 
 $1$& 
 $-1$
 & $A,E$\\
\cline{4-6}
&&& $-1$& $1$ &N/A\\ 
\cline{4-6}
       \hline

         \multirow{3}{*}{$T_h$}& \multirow{3}{*}{$\mbz_2^2$}& \multirow{3}{*}{($\omega(C_2,C_2),\omega(i,i)$)}  & 
 $(1,1)$& 
 $(-1,1)$
 & $A_g,E_g$\\
\cline{4-6}
&&& $(1,-1)$ & $(-1,-1)$&$A_u,E_u$\\ 
\cline{4-6}
&&&other cases&other cases&N/A\\ 
\cline{4-6}
       \hline 

       \multirow{3}{*}{$O$}& \multirow{3}{*}{$\mbz_2^2$}& \multirow{3}{*}{($\omega(C_2,C_2),\omega(C_2',C_2')$)}  & 
 $(1,1)$& 
 $(-1,-1)$
 & $A_1$\\
\cline{4-6}
&&& $(1,-1)$& $(-1,1)$ &$A_2$\\ 
\cline{4-6}
&&&other cases&other cases&N/A\\ 
\cline{4-6}
       \hline

    \multirow{3}{*}{$T_d$}& \multirow{3}{*}{$\mbz_2^2$}& \multirow{3}{*}{($\omega(C_2,C_2),\omega(\sigma_d,\sigma_d)$)}  & 
 $(1,1)$ & 
 $(-1,-1)$
 & $A_1$\\
\cline{4-6}
&&& $(1,-1)$& 
 $(-1,1)$ &$A_2$\\ 
\cline{4-6}
&&&other cases&other cases&N/A\\ 
\cline{4-6}
       \hline 

 \multirow{5}{*}{$O_h$}& \multirow{5}{*}{$\mbz_2^4$}& \multirow{5}{*}{($\omega(C_2,C_2),\omega(C_2',C_2'),\omega(i,i),\frac{\omega(i,C_2')}{\omega(C_2',i)}$)}  & 
 $(1,1,1,1)$& 
 $(-1,-1,1,1)$
 & $A_{1g}$\\
\cline{4-6}
&&&  $(1,1,-1,1)$&  $(-1,-1,-1,1)$ &$A_{1u}$\\ 
\cline{4-6}
&&&  $(1,-1,1,1)$&  $(-1,1,1,1)$ &$A_{2g}$\\ 
\cline{4-6}
&&&  $(1,-1,-1,1)$&  $(-1,1,-1,1)$ &$A_{2u}$\\ 
\cline{4-6}
&&&other cases&other cases&N/A\\ 
\cline{4-6}
       \hline 
\end{longtable}

\bibliographystyle{apsrev}
\bibliography{psg}

\end{document}